\documentclass[aps,prd,preprint,superscriptaddress,nofootinbib]{revtex4-1}

\usepackage{amsmath}
\usepackage{amssymb}
\usepackage{graphicx}
\usepackage{dcolumn}
\usepackage{bm}
\usepackage{hyperref}
\usepackage{epsfig}
\usepackage{mathrsfs}
\usepackage{slashed}

\usepackage{setspace}

\begin{document}

\preprint{TUM-EFT 176/23}

\title{Gauge invariance of radiative jet functions in the
  position-space formulation of SCET}



\author{Geoffrey~T.~Bodwin}
\email[]{gtb@anl.gov}
\affiliation{High Energy Physics Division, Argonne National Laboratory,
Argonne, Illinois 60439, USA}
\author{June-Haak~Ee}
\email[]{june\_haak\_ee@fudan.edu.cn}
\affiliation{Key Laboratory of Nuclear Physics and Ion-beam Application (MOE) and Institute of Modern Physics, Fudan University, Shanghai 200433, China}
\author{Daekyoung Kang}
\email[]{dkang@fudan.edu.cn}
\affiliation{Key Laboratory of Nuclear Physics and Ion-beam Application (MOE) and Institute of Modern Physics, Fudan University, Shanghai 200433, China}
\author{Xiang-Peng~Wang}
\email[]{xiangpeng.wang@tum.de}
\affiliation{Technical University of Munich, TUM School of Natural Sciences, Department of Physics T30f, James-Franck-Straße 1, 85748 Garching, Germany}



\date{\today}

\begin{abstract}
In subleading powers of soft-collinear effective theory (SCET), the
Lagrangian contains couplings between soft quarks and hard-collinear
quarks. Matrix elements of the hard-collinear parts of these couplings
are radiative jet functions. In the position-space formulation of SCET,
the Lagrangians are constructed from operators that appear to be gauge
invariant.  Nevertheless, we find violations of gauge invariance arise
    in the hard-collinear sector because gauge transformations can shift
    the momentum of a hard-collinear quark field from the hard-collinear
    sector to the soft sector, where the hard-collinear fields, by
    definition, have no support.  The violations of gauge invariance are
    manifested in perturbation theory in the hard-collinear sector
    through the absence of certain Feynman diagrams that would be
    present in full QCD.  A consequence of the absence of these diagrams
    is that the radiative jet functions that follow directly from the
    position-space Lagrangians are not gauge invariant, and we
    demonstrate this through explicit calculations in lower-order
    perturbation theory.  We obtain gauge-invariant Lagrangians by
    adding to existing position-space Lagrangians terms that are
    proportional to the soft-quark equation of motion. These
    gauge-invariant Lagrangians are valid for nonzero, as well as zero,
    quark masses. We also remark briefly on the gauge invariance of
    certain Lagrangians that have been constructed in the label-momentum
    formulation of SCET.
\end{abstract}


\maketitle




\section{Introduction}

In soft-collinear effective theory (SCET) \cite{Bauer:2000ew,
  Bauer:2000yr, Bauer:2001ct, Bauer:2001yt, Bauer:2002nz}, couplings
between a quark that carries a soft momentum and a quark that carries a
hard-collinear momentum appear in subleading powers of the SCET
expansion parameter $\lambda$ \cite{Beneke:2002ph,
  Beneke:2002ni,Pirjol:2002km,Bauer:2003mga}.  Matrix elements that
contain the hard-collinear parts of these couplings are called
``radiative jet functions,'' and they appear in factorization
theorems for exclusive processes at subleading power in $\lambda$.
(See, for example, Refs.~\cite{Moult:2019mog,Liu:2019oav,Beneke:2019oqx}.)

In general, radiative jet functions are written in terms of operators in
which hard-collinear quark and antiquark fields are accompanied by
Wilson-line factors, and all derivatives are covariant derivatives.  We
call such operators ``ostensibly gauge-invariant operators.''  If one
replaces the hard-collinear quark and antiquark fields with full QCD
fields multiplied by collinear projectors, then the ostensibly
gauge-invariant operators are truly gauge invariant.  However, in SCET,
the hard-collinear quark and antiquark fields must carry hard-collinear
momenta, not soft momenta.\footnote{\setstretch{0.8}The requirement that hard-collinear
quark and gluon fields carry hard-collinear momenta is essential in
working out SCET power counting and in achieving a factorization of the
hard-collinear sector of the effective theory from the other sectors at
the Lagrangian level. The assumption that the hard-collinear quark and
gluon fields carry hard-collinear momenta is used explicitly in
Refs.~\cite{Beneke:2002ph, Beneke:2002ni} in constructing the
Lagrangians in those papers.}  This requirement can lead to violations
    of hard-collinear gauge invariance because hard-collinear gauge
    transformations multiply quark fields by a phase that, in momentum
    space, can shift the quark-field momentum from the hard-collinear
    region to the soft region, where the hard-collinear fields have no
    support.  As we will see, this phenomenon is manifested
    diagrammatically by the absence of certain Feynman diagrams in the
    hard-collinear sector that would be present in full QCD. These
    ``missing diagrams'' can lead to violations of gauge invariance in
    the hard-collinear sector.

We carry out an analysis of the Lagrangians that appear in the
position-space formulation of SCET in Refs.~\cite{Beneke:2002ph,
  Beneke:2002ni}, which we refer to as Beneke-Chapovsky-Diehl-Feldmann
    (BCDF).\footnote{\setstretch{0.8}The Lagrangians in Eq.~(A.1) of
        Ref.~\cite{Beneke:2019oqx} are the BCDF Lagrangians, but
        expressed in terms of gauge-invariant building blocks.  In this
        paper, we carry out our analyses in terms of the original BCDF
        forms of the Lagrangians.} We demonstrate, through the use of
    Ward identities, that the BCDF Lagrangians lead to violations of
gauge invariance in the hard-collinear sector. The violations occur in
order $\lambda^2$, but not in order $\lambda^1$. We find that the
violations of gauge invariance can be removed by making use of the
soft-quark equation of motion.  Therefore, the BCDF Lagrangians lead
to gauge-invariant $S$-matrix elements \cite{Arzt:1993gz}.
However, off-shell quantities, such as radiative jet functions can be
gauge noninvariant. We use the Lagrangians in
    Refs.~\cite{Beneke:2002ph, Beneke:2002ni} directly to construct
    radiative jet functions. That is, we define the radiative jet
    functions as time-ordered matrix elements of the
    hard-collinear-operator factors in the Lagrangians. We find, through
    explicit calculations at the lowest nontrivial order in perturbation
    theory, that the resulting radiative jet functions are not gauge
    invariant.
 
We modify the BCDF Lagrangians to obtain gauge-invariant Lagrangians
that describe the couplings of a soft quark to a hard-collinear quark by
applying the soft-quark equation of motion and by making use of the
    Bauer-Pirjol-Stewart (BPS) field redefinition in Ref.~\cite{Bauer:2001yt} to factor
    minus-polarized soft gluons from the hard-collinear subdiagram.  We
    use the concept of missing diagrams to argue that the modified
    order-$\lambda^2$ Lagrangians, as well as the order-$\lambda^1$
    Lagrangian, are gauge invariant to all orders in perturbation
    theory.  We also demonstrate the gauge invariance by using the
    modified order-$\lambda^2$ Lagrangians to construct radiative
    jet functions and computing the radiative jet functions in the
    Feynman gauge and the light-cone gauge at the lowest nontrivial
    order in perturbation theory.

In the label-momentum formulation of SCET \cite{Bauer:2000yr}, the
Lagrangians that describe the interactions of soft quarks with
hard-collinear quarks are also constructed from ostensibly
gauge-invariant operators \cite{Bauer:2001ct}.  We find that the
label-momentum Lagrangians in Refs.~\cite{Pirjol:2002km,Bauer:2003mga}
evade the gauge invariance issue that we identify in this paper and that
the corresponding operators are truly gauge invariant.

The remainder of this paper is organized as follows.  In
Sec.~\ref{sec:Preliminaries}, we establish the notations and conventions
that we use throughout this paper.  In Sec.~\ref{sec:ward}, we present
the Lagrangians of Refs.~\cite{Beneke:2002ph, Beneke:2002ni}, discuss
the associated Ward identities in lowest-order perturbation theory, and
identify the sources of violations of gauge invariance as the
``missing diagrams.'' In Sec.~\ref{sec:radiative-BCDF}, we use the
BCDF Lagrangians to compute radiative jet functions in the Feynman
gauge and in the light-cone gauge at lowest order in perturbation
theory, and we show that these two gauges give different results,
verifying the violation of gauge invariance. In
Sec.~\ref{sec:gauge-inv-lagrangian} we modify the BCDF Lagrangians at
    relative order $\lambda^2$ to obtain gauge-invariant Lagrangians
that connect a soft quark to a hard-collinear quark. In
Sec.~\ref{sec:all-orders-gauge-inv}, we argue that the order-$\lambda^1$
Lagrangian and the modified order-$\lambda^2$ Lagrangians are gauge
invariant to all orders in perturbation theory. We construct radiative
jet functions that follow from the modified gauge-invariant
Lagrangians in Sec.~\ref{sec:radiative-new}, and we calculate these
radiative jet functions in the Feynman gauge and in the light-cone gauge
in lowest-order perturbation theory, verifying that the radiative jet
functions are invariant with respect to these gauge choices. In
Sec.~\ref{sec:label-momentum}, we observe that certain versions of the
label-momentum formulation of SCET evade the gauge-invariance problem
that we identify in this paper.  Finally, we summarize and discuss our
results in Sec.~\ref{sec:discussion}.

\section{Preliminaries}
\label{sec:Preliminaries}
In this section, we establish the notations and conventions that we use
throughout this paper.

We decompose an arbitrary vector in terms of the two lightlike vectors, 
$n$ and $\bar{n}$, as follows:
\begin{eqnarray}
r^\mu = r^- \frac{\bar{n}^\mu}{2}
+r^+ \frac{n^\mu}{2}
+r_\perp^\mu,
\end{eqnarray}
where 
\begin{eqnarray}
\label{eq:light-front-coordinate-convention}
r^+ = \bar{n}\cdot r,
\quad
r^- = n\cdot r,
\quad
r_\perp^\mu = r^\mu 
-r^- \frac{\bar{n}^\mu}{2}
-r^+ \frac{n^\mu}{2},
\end{eqnarray}
with $n^2 = \bar{n}^2 = 0$ and $n\cdot \bar{n}=2$. 
$n$ and $\bar{n}$ are the lightlike unit vectors 
along the $z$ axis:
\begin{equation}
\label{eq:n-nbar-def}
n^\mu = (1,0,0,1),
\quad
\bar{n}^\mu = (1,0,0,-1).
\end{equation}
The perpendicular momentum $r_\perp^\mu$ satisfies $n\cdot r_\perp =
\bar{n}\cdot r_\perp =0$. Here, and throughout this paper, we use the
notation
\begin{equation}
r_\perp^2 = -\bm{r}_\perp^2,
\end{equation}
where $\bm{r}_\perp$ is a $(D-2)$-dimensional Euclidean vector.

It is convenient to consider the case of a quark with mass $m\gg
\Lambda_{\rm QCD}$.  Then, a soft momentum $r_s$ on the quark line has
the scaling behavior
\begin{eqnarray}
\label{eq:soft-ell-power}
r_s^+ \sim Q\lambda^2,
\quad
r_s^- \sim Q\lambda^2,
\quad
r_{s\perp}\sim Q\lambda^2,
\end{eqnarray}
where \footnote{\setstretch{0.8}Our general arguments and specific examples are also
  valid for massless quarks for those cases that do not involve an
  interaction between a soft-quark and a hard-collinear quark that is
  proportional to $m$.  In the massless case, the SCET scaling parameter
  is $\lambda=(\Lambda_{\rm QCD}/Q)^{1/2}$.}
\begin{equation}
\label{eq:power-count-lam}
\lambda = \sqrt{\frac{m}{Q}},
\end{equation}
and $Q$ is the hard scale of the process. A collinear momentum $r_c$ on the
    quark line along the $n$ direction has the scaling behavior 
\begin{eqnarray}
r_c^+ \sim Q,
\quad
r_c^- \sim Q\lambda^4,
\quad
r_{c\perp}\sim Q\lambda^2.
\end{eqnarray}
Since a radiative jet function
carries a soft momentum combined with a collinear momentum, the
resulting hard-collinear momentum $r_{hc}$, taken to be along the $n$
direction, has the scaling behavior 
\footnote{\setstretch{0.8}The scaling of $r_{hc\perp}$ is chosen so that $r_{hc
        \perp}^2 \sim r_{hc}^+r_{hc}^-$.}
\begin{eqnarray}
\label{eq:n-hard-collinear-mom}
r_{hc}^+ \sim Q,
\quad
r_{hc}^- \sim Q\lambda^2,
\quad
r_{hc\perp}\sim Q\lambda.
\end{eqnarray}
Note that the hard-collinear and soft momenta have different
virtualities in $\lambda$: $r_{hc}^2 \sim Q^2\lambda^2$, and $r_s^2\sim Q^2
\lambda^4$.

The $n$-hard-collinear Dirac field $\psi_n$ can be decomposed into 
large- and small-component collinear fields by applying
collinear projectors $P_n$ and $P_{\bar{n}}$ onto $\psi_n$:
\begin{subequations}%
\label{eq:xi-projectors}%
\begin{eqnarray}
\label{eq:xi-n}%
\xi_n&=&P_n\psi_n,\\
\eta_{n}&=&P_{\bar n}\psi_{n},
\end{eqnarray}
\end{subequations}
where
\begin{equation}
P_n = \frac{\slashed{n}\slashed{\bar{n}}}{4},
\quad
P_{\bar{n}}
=
\frac{\slashed{\bar{n}}\slashed{n}}{4}.
\end{equation}
We make use of the following additional notations for SCET fields:
    $q_s$ is a soft-quark field, $G_n^\mu$ is an $n$-hard-collinear-gluon field, and $G_s^\mu$ is a soft-gluon field.  $q_s$, $\eta_n$
    and $\xi_n$ have scaling of order $\lambda^3$,
        $\lambda^2$ and $\lambda$, respectively. Each component of
  the field $G_n^\mu$ has the same scaling as the
  $n$-hard-collinear momentum in Eq.~(\ref{eq:n-hard-collinear-mom}),
  and each component of the field $G_s^\mu$ has the same scaling
  as the soft momentum in Eq.~(\ref{eq:soft-ell-power}).  $g_s =
  \sqrt{4\pi \alpha_s}$ is the strong coupling.

We define the $n$-hard-collinear Wilson line as
\begin{eqnarray}
\label{def:collinear-Wilson}
W_n(x) &=& P\exp\left[ig_s\int_{-\infty}^0 ds\, \bar{n}\cdot G_n(x+s\bar{n})\right],
\end{eqnarray}
and we define the covariant derivatives as
\begin{eqnarray}
\label{def:covariant-derivative}
D^\mu &=& \partial^\mu - i g_s G_n^\mu(x) - ig_s n\cdot G_s(x^+)
\frac{\bar{n}^\mu}{2},
\nonumber \\
D_n^\mu &=& \partial^\mu - i g_s G_n^\mu(x),
\nonumber \\
D_s^\mu &=& \partial^\mu - ig_s G_s^\mu(x).
\end{eqnarray}

\section{Ward Identities of the BCDF SCET Lagrangians \label{sec:ward}}
\subsection{BCDF Lagrangians\label{sec:BCDF-Lagrangians}}

The effective Lagrangians that describe an interaction between a soft
quark $q_s$ and an $n$-hard-collinear quark $\xi_n$ in the SCET
formulation Refs.~\cite{Beneke:2002ph, Beneke:2002ni} are given by
\begin{subequations}%
\label{eq:collinear-soft-all}%
\begin{eqnarray}
\label{eq:collinear-soft-all-1}%
{\cal L}_{1}^{\rm BCDF}(x)
&=&
\bar{q}_s(x^+)\left(W_n^\dagger i\slashed{D}_{n\perp} \xi_n\right)(x)
+\textrm{H.c.},
\\
\label{eq:collinear-soft-all-2a}%
{\cal L}_{2a}^{\rm BCDF}(x)
&=&
\bar{q}_s(x^+)
\left\{
W_n^\dagger
\left(
in\cdot D
+
i\slashed{D}_{n\perp}
\frac{1}{i\bar{n}\cdot D_n} i\slashed{D}_{n\perp}
\right)
\frac{\slashed{\bar{n}}}{2}
\xi_n
\right\}(x)
+\textrm{H.c.},
\\
\label{eq:collinear-soft-all-2b}%
{\cal L}_{2b}^{\rm BCDF}(x)
&=&
\left[
\bar{q}_s
(-i\overleftarrow{D}_{s\perp}^\rho)
\right]
(x^+)
\left(
ix_{\perp\rho}
W_n^\dagger i \slashed{D}_{n\perp}
\xi_{n}
\right)
(x)
+\textrm{H.c.},
\\
\label{eq:collinear-soft-all-2m}%
{\cal L}_{2m}^{\rm BCDF}(x)
&=&
\bar q_s(x^+)
\left(
-
m
W_n^{\dagger}
\xi_n
\right)(x)
+\textrm{H.c.},
\end{eqnarray}
\end{subequations}%
 where H.c.\ denotes the Hermitian conjugate, and the subscripts 1 and 2
 indicate the order in $\lambda$ of these expressions.  These
     Lagrangians generalize slightly those in Refs.~\cite{Beneke:2002ph,
       Beneke:2002ni}, in that they contain a nonzero quark mass.
     However, we refer to them as the BCDF Lagrangians.  In deriving
 these expressions, we have started in full QCD, with a quark with mass
 $m$, and we have followed the steps that are given in
 Refs.~\cite{Beneke:2002ph, Beneke:2002ni} for the massless case.  The
 detailed derivation of ${\cal L}_{2m}^{\rm BCDF}$ is given in
 Appendix~\ref{app:mass-lag}.

References~\cite{Beneke:2002ph, Beneke:2002ni} also contain the $\bar
\xi_{n} \ldots \xi_{n}$, $\bar q_{s} \ldots q_{s}$, and pure gauge-field
SCET Lagrangians.  The modifications of these Lagrangians for the
massive case at orders $\lambda^0$, $\lambda^1$, and $\lambda^2$ are
also shown in Appendix~\ref{app:mass-lag}. In this paper, we do not use
these Lagrangians explicitly. Instead, we employ an equivalent, but
simpler, procedure: we replace the fields $\xi_{n}$ and $\bar \xi_{n}$
with hard-collinear Dirac fields, using the expressions in
Eq.~(\ref{eq:xi-projectors}); we use full-QCD Feynman rules, with the
projectors $P_n$ and $P_{\bar n}$; and we expand to the desired order in
$\lambda$.

The power counting in $\lambda$ in Eq.~(\ref{eq:collinear-soft-all})
follows from the fact that, if we integrate the interaction Lagrangians
in Eq.~(\ref{eq:collinear-soft-all}) over $d^4 x$, then the integration
region is of order $\lambda^{-4}$. In ${\cal L}_{2b}^{\rm BCDF}(x)$,
the factor $ix_{\perp\rho}$ should be counted as $O(\lambda^{-1})$
because it scales as the inverse of the transverse momentum that flows
into the hard-collinear subdiagram.

Note that, in Eq.~(\ref{eq:collinear-soft-all}), the soft-quark field
    $q_s$ depends only on $x^+$. That is, the soft-quark field has been
    multipole expanded in the minus and transverse components of its
    argument in order to obtain a definite scaling in $\lambda$
    \cite{Beneke:2002ph}.  Square brackets indicate that a derivative
    acts only inside the brackets and that soft fields are evaluated at
    $x^+$ after the derivative is taken.

The factor $ix_{\perp\rho}$ becomes, in momentum space, a derivative
with respect to the transverse component of the soft momentum.  In
    momentum space, the multipole expansion of $q_s$ implies that the
plus and transverse components of the soft momentum are ultimately set
to zero, but only after derivatives with respect to the soft momentum
have been taken.

At this stage, minus-polarized soft gluons still attach to the
hard-collinear subdiagram.  These attachments can be factored into
Wilson lines by making use of the Grammer-Yennie approximation
\cite{Grammer:1973db}, followed by the application of perturbative Ward
identities \cite{Collins:1988ig}.  Equivalently, the decoupling can be
achieved by making use of the BPS collinear-field redefinitions
\cite{Bauer:2001yt}:
\begin{eqnarray}
\label{eq:BPS-redef}
\xi_n(x) \to S_n(x^+) \xi_n(x),
\quad
G_n^\mu(x) \to S_n(x^+) G_n^\mu(x) S_n^\dagger(x^+),
\end{eqnarray}
where $S_n$ is the soft Wilson line, which is defined by
\begin{eqnarray}
\label{def:Wilson-line}%
S_{n}(x)&=& P~ \text{exp}\left[ig_{s}\int_{-\infty}^{0}dt\, n\cdot G_{s}
(x+tn)\right].
\end{eqnarray}
The net effect of the BPS field redefinitions is to make the simple
    replacements $\bar{q}_s(x^+)\to \bar{q}_s(x^+) S_n(x^+)$,
    $[\bar{q}_s(-i\overleftarrow{D}_{s\perp}^\rho)](x^+) \to
    [\bar{q}_s(-i\overleftarrow{D}_{s\perp}^\rho)](x^+)S_n(x^+)$, and
    $in\cdot D \to in\cdot D_n$ in Eq.~(\ref{eq:collinear-soft-all}).
    Note that the arguments of the soft Wilson lines have been multipole
    expanded to lowest order in $\lambda$, and, so, it is still true
    that only the minus component of the soft momentum enters the
    momentum-space expressions that derive from the BPS-transformed
    Lagrangians.

\subsection{Gauge invariance and Ward identities}
\label{sec:ward-original}

At first sight, the hard-collinear parts of the interactions in
Eq.~(\ref{eq:collinear-soft-all}) would appear to be gauge invariant
with respect to gauge transformation of the hard-collinear fields. We
can use Eq.~(\ref{eq:xi-n}) to replace $\xi_n$ with $\psi_{n}$.  Then,
if we could replace $\psi_n$ with the ordinary Dirac field $\psi$, the
hard-collinear parts of the interactions in
Eq.~(\ref{eq:collinear-soft-all}) would be gauge-invariant full-QCD
operators. However, as we will see, the restriction that $\psi_{n}$ (and
$\xi_{n}$) carry $n$-hard-collinear momenta implies that certain
full-QCD diagrams are missing in hard-collinear functions involving
these operators and that, consequently, the hard-collinear functions are
not gauge invariant.

\begin{figure}
\centering
\includegraphics[scale=0.6]{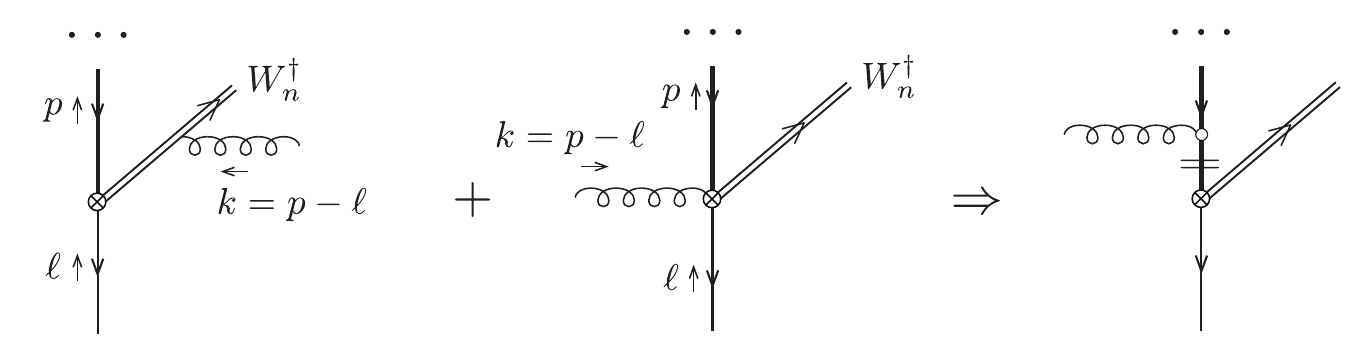}
\caption{\label{fig:A1} \setstretch{1.5}Feynman diagrams showing a Ward identity for the
  operators in Eq.~(\ref{eq:collinear-soft-all}).  The Ward identity is
  for the case in which a gluon attaches to the Wilson line
  ($W_n^\dagger$) or to the crossed circle $(\otimes)$.  The thick,
  solid line denotes a quark line that has a hard-collinear momentum
  ($-p$), and the solid line denotes a quark line that has a soft
  momentum ($-\ell$).  The diagram on the right side of the arrow shows
  the contribution of the term $k^\mu$ in the gauge transformation in
  Eq.~(\ref{eq:gauge-transform}).  The short, double lines across a
  propagator indicate that the propagator has been canceled.}
\end{figure}

\begin{figure}
\centering
\includegraphics[width=0.7\columnwidth]{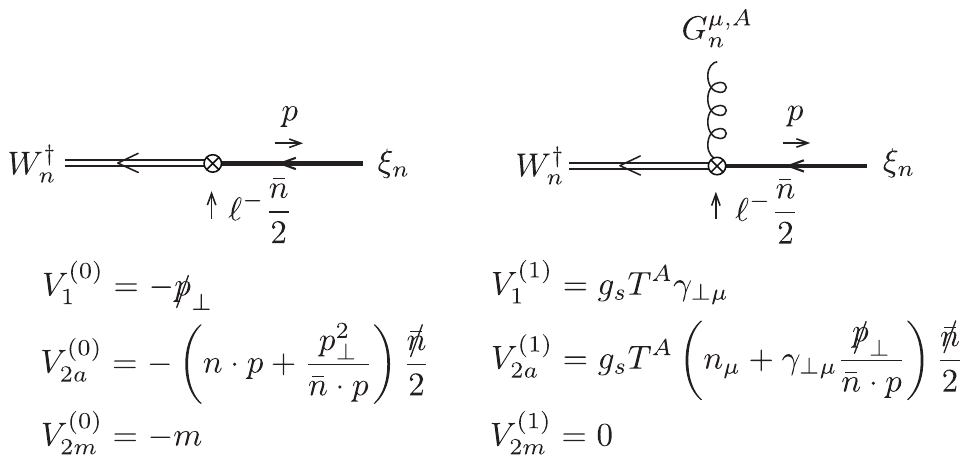}
\caption{\label{fig:feyn-Ajpsi} \setstretch{1.5}The Feynman rules for the crossed
  circles in Fig.~\ref{fig:A1} for the Lagrangians in
  Eqs.~(\ref{eq:collinear-soft-all-1}),
  (\ref{eq:collinear-soft-all-2a}), and
  (\ref{eq:collinear-soft-all-2m}).  $V_{1}$, $V_{2a}$, and $V_{2m}$ are
  the crossed-circle contributions that arise from ${\cal L}_{1}^{\rm
    BCDF}$, ${\cal L}_{2a}^{\rm BCDF}$, and ${\cal L}_{2m}^{\rm BCDF}$,
  respectively.  The superscripts $(0)$ and $(1)$ denote the
  order-$g_s^0$ and $g_s^1$ contributions of the crossed circle,
  respectively.}
\end{figure}

\begin{figure}
\centering
\includegraphics[width=0.7\columnwidth]{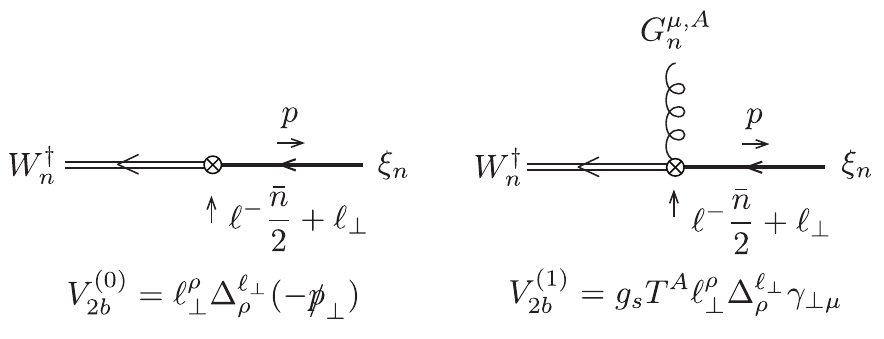}
\caption{\label{fig:feyn-Ajpsi2} \setstretch{1.5}The Feynman rules for the crossed
  circles in Fig.~\ref{fig:A1} for the Lagrangians in
  Eq.~(\ref{eq:collinear-soft-all-2b}).  $V_{2b}$ describes the
  crossed-circle contribution that arises from ${\cal L}_{2b}^{\rm
    BCDF}$.  The superscripts $(0)$ and $(1)$ denote the order-$g_s^0$
  and order-$g_s^1$ contributions of the crossed circle, respectively.
  The definition of the operator $\Delta_\rho^{\ell_\perp}$ is given in
  Eq.~(\ref{def:del-ellperp-op}). }
\end{figure}

In order to investigate the gauge invariance of the $O(\lambda^2)$
SCET Lagrangians in Eq.~(\ref{eq:collinear-soft-all}), let us consider
the interactions that follow from these Lagrangians in order
$g_s$. The corresponding Feynman diagrams are shown on the left
sides of Fig.~\ref{fig:A1}.
 We imagine that these explicit interactions are embedded in a radiative
 jet function, whose remaining factors are indicated by an ellipsis in
 the figures and in the corresponding equations.  In these figures, the
 quark momentum $-p$ is an $n$-hard-collinear momentum, the quark
 momentum $-\ell$ is a soft momentum, and the gluon momentum $k=p-\ell$
 is an $n$-hard-collinear momentum.  The crossed circles in
 Fig.~\ref{fig:A1} arise from the covariant-derivative and $x_\perp$
 factors in the Lagrangians.  Their Feynman rules are given in
 Figs.~\ref{fig:feyn-Ajpsi} and \ref{fig:feyn-Ajpsi2}. The order-$g_s^0$
 contributions to the crossed circles can appear only in conjunction
 with one or more hard-collinear gluons that attach to the Wilson line.
 Otherwise, hard-collinear momentum would not be conserved at
 the crossed circles.
In deriving these Feynman rules, we make use of the identity 
\begin{eqnarray}
\label{eq:identity-ix}
i x_{\perp\rho}
&=&
\int d^{D-2}\ell_\perp
\frac{\partial}{\partial \ell_\perp^\rho}
\left[
\delta^{D-2}(\ell_\perp)
\right]
e^{-i\ell_\perp\cdot x_\perp},
\end{eqnarray}
and we define the operator
\begin{eqnarray}
\label{def:del-ellperp-op}
\Delta^{\ell_\perp}_{\rho}
\equiv
\int d^{D-2}\ell_\perp
\delta^{D-2}(\ell_\perp)
\frac{\partial}{\partial \ell_\perp^\rho},
\end{eqnarray}
which picks up the coefficient of $\ell_\perp^\rho$. 

The amplitudes for ${\cal L}_{1}^{\rm BCDF}$, ${\cal L}_{2a}^{\rm
  BCDF}$, ${\cal L}_{2b}^{\rm BCDF}$, and ${\cal L}_{2m}^{\rm
  BCDF}$, which correspond to the diagram in Fig.~\ref{fig:A1}, are
\begin{eqnarray}
\label{eq:ABBm}
A_{1}
&=&
\epsilon^\mu(k)
\left[\frac{(-ig_s)(i) \bar{n}_\mu}{\bar{n}\cdot k
+i\varepsilon} (-\slashed{p}_\perp)
+ g_s\gamma_{\perp\mu} \right]P_n
\ldots,
\nonumber \\
A_{2a}
&=&
\epsilon^\mu(k) \left[\frac{(-ig_s)(i) \bar{n}_\mu}{\bar{n}\cdot k
+i\varepsilon}  
\left(
-n\cdot p-\frac{{p}_\perp^2}{\bar{n}\cdot p}
\right)
+ g_s
\left(
n_\mu
+
\gamma_{\perp\mu}
\frac{\slashed{p}_\perp}{\bar{n}\cdot p}
\right)
\right]\frac{\slashed{\bar{n}}}{2} P_n
\ldots,
\nonumber \\
A_{2b}
&=&
\epsilon^\mu(k) 
\left[\frac{(-ig_s)(i) \bar{n}_\mu}{\bar{n}\cdot k
+i\varepsilon}
\ell_{\perp}^\rho
\Delta^{\ell_\perp}_{\rho}
(-\slashed{p}_\perp)
+
g_s
\ell_{\perp}^\rho
\Delta^{\ell_\perp}_{\rho}
\gamma_{\perp\mu}
\right]
P_n\ldots,
\nonumber \\
A_{2m}
&=& 
\epsilon^\mu(k) \frac{(-ig_s)(i) \bar{n}_\mu}{\bar{n}\cdot k
+i\varepsilon} (-m)P_n\ldots,
\end{eqnarray}
respectively. Here, $\epsilon^\mu(k)$ is the polarization of a
hard-collinear gluon with momentum $k$. Note that, in $A_1$, $A_{2a}$,
and $A_{2m}$, $p^+=k^+$ and $p_\perp=k_\perp$, owing to the multipole
expansion of the soft-quark field. In $A_{2b}$, $p^+=k^+$, but $p_\perp$
is set equal to $k_\perp$ only after differentiation with respect to
  $\ell_\perp$, in accordance with Eq.~(\ref{def:del-ellperp-op}).

We can work out the Ward identities for these amplitudes by carrying out
    a gauge transformation on the gluon field, which, at the lowest
    order in $g_s$, simply shifts the gluon polarization vector
    $\epsilon^\mu(k)$ by an amount that is proportional to the gluon
    momentum $k$:
\begin{eqnarray}
\epsilon^\mu(k)&\to & \epsilon^\mu(k)+\beta k^\mu.
\label{eq:gauge-transform}
\end{eqnarray}
We drop the constant of proportionality $\beta$ in subsequent discussions.
The gauge term $k^\mu$ gives the following contributions to the
amplitudes $A_{1}$, $A_{2a}$, and $A_{2m}$ in Eq.~(\ref{eq:ABBm}):
\begin{subequations}%
\label{eq:ABBm-gauge}%
\begin{eqnarray}
A_{1}^\textrm{gauge}
&=&
g_s(-\slashed{p}_\perp +\slashed{k}_\perp)P_n \ldots
\nonumber\\
\label{eq:ABBm-gauge-a}%
&=&
0,
\\
\label{eq:ABBm-gauge-b}%
A_{2a}^\textrm{gauge}&=&
g_s
\left[
\left(
-n\cdot p-\frac{{p}_\perp^2}{\bar{n}\cdot p}
\right)
+
\left(
n\cdot k
+
\frac{\slashed{k}_\perp
\slashed{p}_\perp}{\bar{n}\cdot p}
\right)
\right]\frac{\slashed{\bar{n}}}{2} P_n
\ldots
\nonumber \\
&=&
-g_s n\cdot \ell\frac{\slashed{\bar{n}}}{2} P_n
\ldots,
\\
\label{eq:ABBm-gauge-c}%
A_{2m}^\textrm{gauge}&=& g_s(-m)P_n\ldots,
\end{eqnarray}
where we have used $k_\perp = p_\perp$, which follows from the multipole
expansion of the soft-quark field $q_{s}$, and $n\cdot k = n\cdot p - n\cdot
\ell$. The gauge term $k^\mu$ gives the following contribution to the
amplitude $A_{2b}$ in Eq.~(\ref{eq:ABBm}):
\begin{eqnarray}
\label{eq:A_2b-gauge}%
A_{2b}^\textrm{gauge}
&=&
g_s
\left[
\ell_{\perp}^\rho
\Delta^{\ell_\perp}_{\rho}
(-\slashed{p}_\perp)
+
\ell_{\perp}^\rho
\Delta^{\ell_\perp}_{\rho}
\slashed{k}_\perp
\right]
P_n\ldots
\nonumber \\
&=&
g_s
\left[
\ell_{\perp}^\rho
\Delta^{\ell_\perp}_{\rho}
(-\slashed{\ell}_\perp)
\right]
P_n\ldots
\nonumber \\
&=&
-g_s
\slashed{\ell}_\perp
P_n\ldots,
\end{eqnarray}
\end{subequations}%
where we have used $k_\perp =p_\perp-\ell_\perp$, owing to the insertion
of the $\ell_\perp$ into the hard-collinear subdiagram, which follows
from the identity in Eq.~(\ref{eq:identity-ix}).  Note that the
contributions in which $\Delta_\rho^{\ell_\perp}$ acts on the ellipsis
(the remainder of the diagram) cancel between the first and second terms
after the first equality in Eq.~(\ref{eq:A_2b-gauge}).

We see that the order-$\lambda^1$ Lagrangian in
Eq.~(\ref{eq:collinear-soft-all}) gives a vanishing contribution to the
Ward identities. That is, it is gauge invariant.  However, each of the
order-$\lambda^2$ Lagrangians in Eq.~(\ref{eq:collinear-soft-all})
produces a nonzero contribution to the Ward identity, {\it i.e.}, a
violation of gauge invariance, and we find that
\begin{eqnarray}
\label{eq:gauge-violating}
A_{2a}^\textrm{gauge}
+
A_{2b}^\textrm{gauge}
+
A_{2m}^\textrm{gauge}
&=&
g_s
\left(
-n\cdot \ell\frac{\slashed{\bar{n}}}{2}
-\slashed{\ell}_\perp
-m
\right)
P_n\ldots
\nonumber \\
&=&
g_s
(-\slashed{\ell} - m)
P_n\ldots, 
\end{eqnarray}
where we have used $\slashed{n} P_n = 0$.  We note that
the violations of gauge invariance are proportional to soft-quark free
equation of motion. This suggests that we can obtain a gauge-invariant
form of the Lagrangian by discarding terms that are proportional to the
soft-quark equation of motion. In Sec.~\ref{sec:gauge-inv-lagrangian},
we will see that this is the case.
\footnote{\setstretch{0.8}Subtractions of terms that are proportional to the soft-quark
equation of motion are used to derive the Lagrangian in
Ref.~\cite{Pirjol:2002km}.}

A complete factorization formula, including both the radiative jet
(hard-collinear) function and the soft function, must be gauge invariant
because it reproduces full QCD to a given accuracy in
$\lambda$. Therefore, we expect gauge invariance to be restored if we
consider the soft function in conjunction with the radiative jet
function.  As we have seen, the order-$\lambda^2$ contributions to the
radiative jet function that violate gauge invariance are proportional to
the inverse of the soft-quark propagator $-\slashed{\ell}-m$. At the
lowest order in $g_s$, the soft function is just the soft-quark
propagator. Hence, soft function is canceled by the gauge-invariance
violating contributions.  Consequently, as we can see from the result
for the radiative jet functions in Secs.~\ref{sec:radiative-BCDF} and
\ref{sec:radiative-new}, all of the poles in the $\ell^-$ complex plane
are in the upper half plane.  We can then close the $\ell^-$ contour of
integration in the lower half plane to obtain a vanishing result for the
gauge-invariance-violating contribution.  As we have mentioned, in
    Sec.~\ref{sec:gauge-inv-lagrangian}, we will use the formal
    procedure of discarding Lagrangian terms that are proportional to
    the soft-quark equation of motion in order to implement gauge
    invariance in the radiative jet function at the integrand level. The
    phenomenon that we see here, namely, the vanishing of contributions
    that are proportional to the soft-quark equation of motion,
    demonstrates the correctness of the formal procedure in lowest-order
    perturbation theory.

\subsection{Missing diagrams}
\begin{figure}
\centering
\includegraphics[scale=0.5]{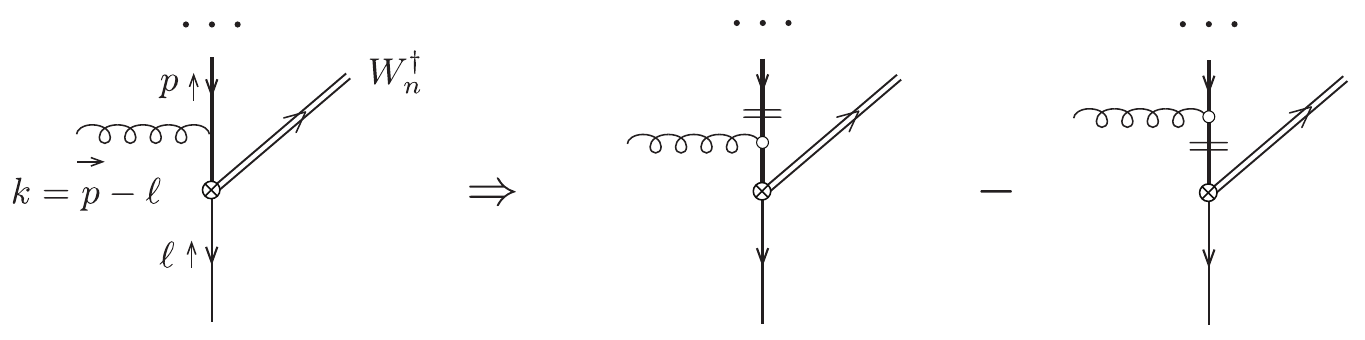}
\caption{\label{fig:A2} \setstretch{1.497}Left side: the Feynman diagram in order
      $g_s$ that is missing from the hard-collinear function. Right
      side: the corresponding Ward-identity contribution. The crossed
      vertex derives from the operators in
      Eq.~(\ref{eq:collinear-soft-all}).}
\end{figure}

The violations of gauge invariance that we have noted arise because
certain diagrams that would be present in full QCD are missing from the
hard-collinear function because they contain a soft-quark
propagator. The diagram of this type that appears in order $g_s$ is
shown on the left side of Fig.~\ref{fig:A2}. Note that, because we are
considering diagrams that contain a soft-quark propagator, momentum
conservation no longer requires that the order-$g_s^0$ factors from the
crossed circles appear in conjunction with hard-collinear gluons that
attach to the Wilson line.

The amplitudes for the diagram in Fig.~\ref{fig:A2} that correspond to
the crossed vertices from
${\cal L}_{1}^{\rm BCDF}$, ${\cal
      L}_{2a}^{\rm BCDF}$, ${\cal L}_{2b}^{\rm BCDF}$, and ${\cal
      L}_{2m}^{\rm BCDF}$ are
\begin{subequations}%
\begin{eqnarray}
\label{eq:ABBm2}%
\label{eq:ABBm2-a}%
A_{1,\textrm{miss}}
&=& 0,
\\
\label{eq:ABBm2-b}%
A_{2a,\textrm{miss}}&=&
\epsilon^\mu(k) (-n\cdot \ell)
\frac{\slashed{\bar{n}}}{2} P_n
\frac{i}{-\slashed{\ell}-m+i\varepsilon}
(ig_s) \gamma_\mu 
\ldots,
\\
\label{eq:ABBm2-c}%
A_{2b,\textrm{miss}}&=&
\epsilon^\mu(k) 
\ell_\perp^\rho \Delta_\rho^{\ell_\perp}
(-\slashed{\ell}_\perp)
P_n
\frac{i}{-\slashed{\ell}-m+i\varepsilon}
(ig_s) \gamma_\mu 
\ldots,
\\
\label{eq:ABBm2-d}%
A_{2m,\textrm{miss}}&=&
\epsilon^\mu(k)
(-m) P_n
\frac{i}{-\slashed{\ell}-m+i\varepsilon}
 (ig_s) \gamma_\mu 
\ldots,
\end{eqnarray}
\end{subequations}%
respectively. Here, in keeping with the multipole expansion of the
soft-quark field $q_s$ in the Lagrangians in
Eq.~(\ref{eq:collinear-soft-all}), we have discarded terms in the
hard-collinear part that are proportional to $\ell_\perp $, except in
$A_{2b,\textrm{miss}}$. In the case of $A_{2b,\textrm{miss}}$,
we set $\ell_\perp$ to zero only after the derivative in
$\Delta_\rho^{\ell_\perp}$ has been taken.  After we carry out the gauge
transformation in Eq.~(\ref{eq:gauge-transform}), we obtain the
following contributions of the gauge term $k^\mu$ to
Eqs.~(\ref{eq:ABBm2}):
\begin{subequations}%
\label{eq:ABBm2-gauge}%
\begin{eqnarray}
\label{eq:order-lambda-gauge}%
A_{1,\textrm{miss}}^\textrm{gauge}
&=&0,
\\
A_{2a,\textrm{miss}}^\textrm{gauge}&=& 
-n\cdot \ell
\frac{\slashed{\bar{n}}}{2}
P_n 
\frac{i}{-\slashed{\ell}-m+i\varepsilon}
(ig_s) \slashed{k}
\ldots
\nonumber\\
&=&
-g_s n\cdot \ell
\frac{\slashed{\bar{n}}}{2}
P_n 
\frac{1}{-\slashed{\ell}-m+i\varepsilon}
[(-\slashed{p}-m)-(-\slashed{\ell}-m)]
\ldots,
\\
A_{2b,\textrm{miss}}^\textrm{gauge}&=&
\ell_\perp^\rho \Delta_\rho^{\ell_\perp}
(-\slashed{\ell}_\perp)
P_n
\frac{i}{-\slashed{\ell}-m+i\varepsilon}
(ig_s) \slashed{k}
\ldots
\nonumber\\
&=&
g_s
\ell_\perp^\rho \Delta_\rho^{\ell_\perp}
(-\slashed{\ell}_\perp)
P_n
\frac{1}{-\slashed{\ell}-m+i\varepsilon}
[(-\slashed{p}-m)-(-\slashed{\ell}-m)]
\ldots,
\\
A_{2m,\textrm{miss}}^\textrm{gauge}&=&
(-m)P_n 
\frac{i}{-\slashed{\ell}-m+i\varepsilon}
(ig_s) \slashed{k}
\ldots
\nonumber\\
&=&
g_s (-m) P_n 
\frac{1}{-\slashed{\ell}-m+i\varepsilon}
[(-\slashed{p}-m)-(-\slashed{\ell}-m)]
\ldots,
\end{eqnarray}
\end{subequations}%
which are represented by the diagram on the right side of
Fig.~\ref{fig:A2}. If the quark line with momentum $-p$ is an external
line, then the first terms in the square brackets in each of the
contributions above vanish on multiplying the quark spinor.  Otherwise,
they cancel contributions that arise from the gauge transformations for
the sum over all attachments of the gluon with momentum $k$ to other
parts of the radiative jet function (the ellipses in Figs.~\ref{fig:A1}
and \ref{fig:A2}).\footnote{\setstretch{0.8}This can be seen by direct application of
the Ward identities for the elementary QCD vertices and the Wilson line
$W_n$ that appears in the radiative jet function.}  The second term in
the square brackets in each of the contributions above cancels the
corresponding gauge terms in Eqs.~(\ref{eq:ABBm-gauge-b}),
(\ref{eq:ABBm-gauge-c}), and (\ref{eq:A_2b-gauge}).  This cancellation
confirms our assertion that the violations of gauge invariance arise
because of missing diagrams involving soft-quark propagators.  Note that
the contribution in Eq.~(\ref{eq:order-lambda-gauge}) that arises from
the order-$\lambda^1$ Lagrangian vanishes, in keeping with the gauge
invariance of that Lagrangian.

\section{Radiative jet function in order $\alpha_s$
from the BCDF Lagrangians \label{sec:radiative-BCDF}} Now let us test
    the gauge invariance of radiative jet functions that are derived from
    the BCDF Lagrangians.  From the BCDF Lagrangians at $O(\lambda^2)$
in Eq.~(\ref{eq:collinear-soft-all}), we can construct the following
radiative jet functions:
\begin{eqnarray}
\label{def:A-3S1-BCDF}
A(\ell^-)
&=&
\int d^D x\,
e^{-i\frac{\ell^- x^+}{2}}
\nonumber \\
&&
\times
\langle {\cal Q}{\cal\bar{Q}}(^3S_1^{[1]},p,p) | 
T\,
\left[
W_n^\dagger
\left(
in\cdot D_n
+
i\slashed{D}_{n\perp}
\frac{1}{i\bar{n}\cdot D_n} i\slashed{D}_{n\perp}
\right)
\frac{\slashed{\bar{n}}}{2}
\xi_n
\right]^{\beta,b}(x)
(\bar{\xi}_nW_n)^{\alpha,a}(0)
|0\rangle,
\nonumber \\
B_{\rho}(\ell^-)
&=&
\int d^{D-2}\ell_{\perp}
\frac{\partial}{\partial\ell_{\perp}^\rho}
\left[
\delta^{D-2}({\ell}_{\perp})
\right]  
\nonumber \\
&&
\times
\int d^D x\,
e^{-i\frac{\ell^- x^+}{2}}
e^{-i\ell_\perp\cdot x_\perp}
\langle {\cal Q}{\cal\bar{Q}}(^3S_1^{[1]},p,p) | 
T\,
\left(W_n^\dagger i\slashed{D}_{n\perp} \xi_n\right)^{\beta,b}(x)
(\bar{\xi}_nW_n)^{\alpha,a}(0)
|0\rangle,
\nonumber \\
M(\ell^-)
&=&
\int d^D x\,
e^{-i\frac{\ell^- x^+}{2}}
\langle {\cal Q}{\cal\bar{Q}}(^3S_1^{[1]},p,p) | 
T\,
\left(-m
W_n^{\dagger}
\xi_n\right)^{\beta,b}(x)
(\bar{\xi}_nW_n)^{\alpha,a}(0)
|0\rangle,
\end{eqnarray}
where $\alpha,\beta$ and $a,b$ are Dirac and color indices,
respectively. In these expressions, the first factors of the SCET
operators arise from the BCDF Lagrangians ${\cal L}_{2a}^{\rm
  BCDF}$, ${\cal L}_{2b}^{\rm BCDF}$, and ${\cal L}_{2m}^{\rm BCDF}$ in
Eq.~(\ref{eq:collinear-soft-all}), respectively, while the second
factors of the SCET operators, namely, $\bar{\xi}_nW_n$, arise
from the hard-collinear part of the hard-to-hard-collinear transition
operator in the amplitude for a hard-scattering process.  The couplings
of minus-polarized soft-gluon fields to hard-collinear fields
have been removed by making use of the BPS field redefinitions
    [Eq.~(\ref{eq:BPS-redef})]. This has the effect of replacing $D$ in
    $A(\ell^-)$ with $D_n$. We have taken the
    vacuum to $\cal{Q}\cal{\bar Q}$ matrix elements of these operators,
    where $\cal{Q}$ and $\cal{\bar Q}$ are massive, on-shell quark
    states. These matrix elements are convenient because, for them, the
    difficulty with gauge invariance appears at the Born level, rather
    than at the loop level.  For definiteness, we take the $\cal{Q}$ and
    the $\cal{\bar Q}$ to be in a spin-triplet, color-singlet state
    with zero relative momentum between the $\cal{Q}$ and the
    $\cal{\bar Q}$. (This is an $S$-wave state.)  Then, we can take both
    the $\cal{Q}$ and the $\cal{\bar Q}$ to have momentum $p$.  We can,
    without loss of generality, choose a frame in which
    ${p}_\perp=0$. Note that, because the momentum $p$ is on the mass
    shell, it satisfies collinear scaling, rather than hard-collinear
    scaling:
\begin{eqnarray}
\label{eq:p-collinear-power}
p^+&\sim& Q\lambda^0, \nonumber\\
\,{p}_\perp&=&0,\nonumber\\
p^-&=&m^2/p^+\sim Q\lambda^4.
\end{eqnarray}
Note also that the Ward-identity arguments of Sec.~\ref{sec:ward},
which were presented for the case of hard-collinear scaling, are also
valid for the case of collinear scaling.

We realize the $\cal{Q}\cal{\bar Q}$ color and spin states through the
application of the standard spin and color projection operators, whose
product is given by
\begin{eqnarray}
\Pi_{^3S_1^{[1]}}^{cd}= -\frac{\slashed{\epsilon}^*(\slashed{p}+m)}{2\sqrt{2}m}
\times
\frac{\delta^{cd}}{\sqrt{N_c}},
\end{eqnarray}
where $\epsilon^*$ is the polarization vector of the external
$^3S_1^{[1]}$ state, which satisfies 
$\epsilon^*\cdot p=0.$

\begin{figure}
\centering
\includegraphics[width=1\columnwidth]{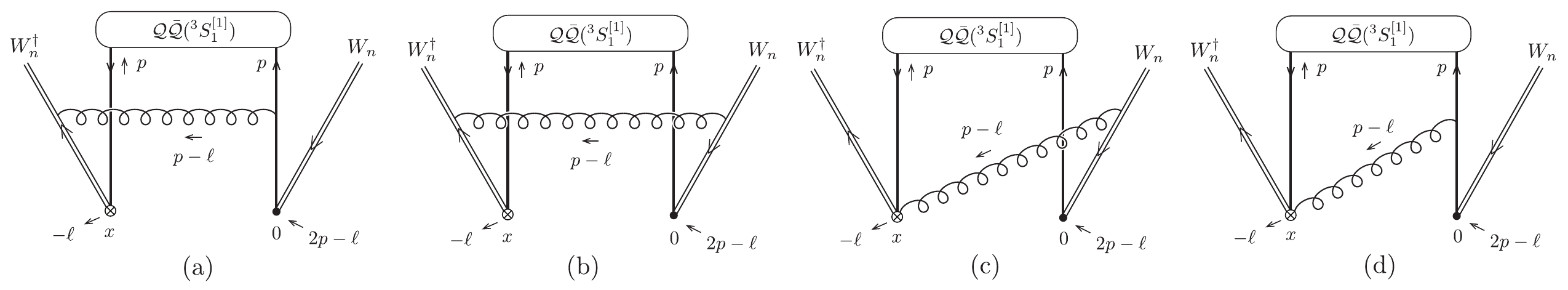}
\caption{\label{fig:AA-LO} \setstretch{1.5}The LO diagrams for the radiative jet
  functions. Note that the soft momentum $\ell^\mu$ flows in through the
  vertex on the soft-quark side at $x$ and flows out through the vertex
  on the hard-quark side at $0$. The Feynman rules for the crossed
  circle vertex are given in Figs.~\ref{fig:feyn-Ajpsi} and
  \ref{fig:feyn-Ajpsi2}.}
\end{figure}

The $O(\alpha_s)$ diagrams for the radiative jet functions are given in
Fig.~\ref{fig:AA-LO}. Note that the gluon must connect the soft-quark side of the diagram, which is to the left of the $\cal{Q}\cal{\bar Q}$
state, to the hard-quark side of the diagram, which is to the right of
the $\cal{Q}\cal{\bar Q}$ state, in order for
hard-collinear-momentum conservation to be satisfied.  The left two
diagrams involve the Feynman rule $V_i^{(0)}$
(Figs.~\ref{fig:feyn-Ajpsi} and \ref{fig:feyn-Ajpsi2}) on their soft-quark sides, and the right two diagrams involve the Feynman rule
$V_i^{(1)}$ (Figs.~\ref{fig:feyn-Ajpsi} and \ref{fig:feyn-Ajpsi2}) on
their soft-quark sides with $i=2a,2b,2m$.

\subsection{Feynman gauge}
\label{sec:feyn-BCDF}

First, let us consider the computations of the radiative jet functions
in the Feynman gauge. 

For the $A$ radiative jet function, only the diagram of
Fig.~\ref{fig:AA-LO}(c) contributes.
The diagram of Fig.~\ref{fig:AA-LO}(b) vanishes because
the Wilson-line vertices give $\bar n \cdot \bar n=0$. The diagrams of
Figs.~\ref{fig:AA-LO}(a) and (d) are power suppressed, as we show in
Appendix~\ref{app:power-suppressed}. 

For the $B$ radiative jet function, only the diagram of
Fig.~\ref{fig:AA-LO}(d) contributes.  The diagram of
Fig.~\ref{fig:AA-LO}(b) vanishes because the Wilson-line vertices give
$\bar n \cdot \bar n=0$.  The diagram of Fig.~\ref{fig:AA-LO}(c)
vanishes because the crossed vertex gives $\bar{n}\cdot
\gamma_{\perp}=0$.  The diagram of Fig.~\ref{fig:AA-LO}(a) is power
suppressed because of the scaling in
Eq.~(\ref{eq:p-collinear-power}).

For the $M$ radiative jet function, only the diagram of
Fig.~\ref{fig:AA-LO}(a) contributes. The diagram of
Fig.~\ref{fig:AA-LO}(b) vanishes because the Wilson-line vertices give
$\bar{n}\cdot \bar{n}=0$, and the diagrams of Figs.~\ref{fig:AA-LO}(c)
and (d) do not contribute because $V_{2m}^{(1)}=0$ 
(Fig.~\ref{fig:feyn-Ajpsi}).

Then, a straightforward calculation in the Feynman gauge yields the
    following Born-level radiative jet functions:
\begin{eqnarray}
\label{eq:rad-Feyn-all}
A_{\textrm{(c),Feynman}}(\ell^-)
&=&
(T^e)^{bc}
(T^e)^{da}
\left\{
g_s
\left(
{n}_\mu
+
\gamma_{\perp\mu}
\frac{\slashed{p}_\perp}{\bar{n}\cdot p}
\right)
\frac{\slashed{\bar{n}}}{2}
P_n
\Pi_{^3S_1^{[1]}}^{cd}
P_{\bar{n}}
\right\}^{\beta\alpha}
\frac{i(ig_s \bar{n}^\mu)}
{-\bar{n}\cdot \left(p-\ell\right)+i\varepsilon}
\frac{-i}{\left(p-\ell\right)^2+i\varepsilon}
\nonumber \\
&=&
-
\frac{ig_s^2C_F\delta^{ba}}{\sqrt{2N_c}mp^+}
\frac{1}
{(-\ell^- +i\varepsilon)}
\left(
\frac{\slashed{\bar{n}}\slashed{n}\slashed{\epsilon}^*}{4}
\right)^{\beta\alpha},
\nonumber \\
B_{\rho\textrm{(d),Feynman}}(\ell^-)
&=&
\int d^{D-2}\ell_{\perp}
\frac{\partial}{\partial\ell_{\perp}^\rho}
\left[
\delta^{D-2}({\ell}_{\perp})
\right]  
(T^e)^{b c}
(T^e)^{da}
\nonumber \\
&&
\times
\left\{
\left(
g_s\gamma_{\perp\mu}
\right)
P_n
\Pi_{^3S_1^{[1]}}^{cd}
(ig_s\gamma^\mu)
\frac{i}{2\slashed{p}-\slashed{\ell}-m+i\varepsilon}
P_{\bar{n}}
\right\}^{\beta\alpha}
\frac{-i}{\left(p-\ell\right)^2+i\varepsilon}
\nonumber \\
&=&0,
\nonumber \\
M_{\textrm{(a),Feynman}}(\ell^-)
&=&
(T^e)^{bc}(T^e)^{da}
\left\{
\left(
-m
\right)
P_n
\Pi_{^3S_1^{[1]}}^{cd}
(ig_s\gamma_\mu)
\frac{i}
{2\slashed{p}-\slashed{\ell}-m+i\varepsilon}
P_{\bar{n}}
\right\}^{\beta\alpha}
\frac{i(-ig_s \bar{n}_\nu)}
{\bar{n}\cdot \left(p-\ell\right)+i\varepsilon}
\nonumber \\
&&
\times
\frac{-ig^{\mu\nu}}{\left(p-\ell\right)^2+i\varepsilon}
\nonumber \\
&=&
-
\frac{ig_s^2C_F\delta^{ba}}{\sqrt{2N_c}p^+}
\frac{1}{(-\ell^-+i\varepsilon)^2}
\left(
\frac{\slashed{n}\slashed{\epsilon}^*}{2}
\right)^{\beta\alpha},
\end{eqnarray}
where we have kept only the leading nonzero power in $\lambda$ in the
last lines of these expressions. In the case of
$B_{\rho\textrm{(d),Feynman}}(\ell^-)$, we have used
\begin{eqnarray}
\left[
\gamma^\perp_\mu P_n
\slashed{\epsilon}^*
(\slashed{p}+m)
\gamma^\mu
(2\slashed{p}-\slashed{\ell}+m)
P_{\bar{n}}
\right]^{\beta\alpha}
&=&
0,
\end{eqnarray}
which is valid through the power in $\lambda$ in which we are interested.

\subsection{Light-cone gauge}
Next, let us consider the radiative jet functions in
the light-cone gauge, which is defined by the gauge condition
\begin{equation}
\bar{n}\cdot G_n\big|_\textrm{light-cone gauge} = 0,
\label{eq:light-cone-def}
\end{equation}
from which it follows that the gluon-propagator polarization sum is
\begin{equation}
-g^{\mu\nu}+\frac{k^{\mu} \bar{n}^\nu+\bar{n}^\mu k^\nu}{k\cdot \bar{n}}.
\label{eq:pol-sum}
\end{equation}
Here, $k$ is the gluon momentum.  In the light-cone gauge, the
$n$-hard-collinear Wilson line [defined in
  Eq.~(\ref{def:collinear-Wilson})] becomes unity, and so the diagrams
involving Wilson lines vanish. (Equivalently, one can see that the
polarization sum in Eq.~(\ref{eq:pol-sum}) vanishes on contraction with
the Wilson-line vertex $-ig\bar{n}_\mu$.)  Consequently, in the
light-cone gauge, only the diagram of Fig.~\ref{fig:AA-LO}(d) can
contribute.

The calculation of the contribution of the diagram of
Fig.~\ref{fig:AA-LO}(d) in the light-cone gauge is straightforward.  We
obtain
\begin{eqnarray}
\label{eq:light-cone-f}
A_{\textrm{(d),light-cone}}(\ell^-)
&=&
(T^e)^{bc}(T^e)^{da}
\left\{
g_s
\left(
{n}_\mu
+
\gamma_{\perp\mu}
\frac{\slashed{p}_\perp}{\bar{n}\cdot p}
\right)
P_n
\Pi_{^3S_1^{[1]}}^{cd}
(ig_s\gamma_\nu)
\frac{i}
{2\slashed{p}-\slashed{\ell}-m+i\varepsilon}
P_{\bar{n}}
\right\}^{\beta\alpha}
\nonumber \\
&&
\times
\frac{i}{\left(p-\ell\right)^2+i\varepsilon}
\left[
-g^{\mu\nu}
+\frac{\left({p}-\ell\right)^\mu \bar{n}^\nu
+\bar{n}^\mu \left({p}-\ell\right)^\nu}
{\left({p}-\ell\right)\cdot \bar{n}}
\right]
\nonumber \\
&=&
-
\frac{2ig_s^2C_F\delta^{ba}}{\sqrt{2N_c}mp^+}
\frac{1}{(-\ell^-+i\varepsilon)}
\left(
\frac{\slashed{\bar{n}}\slashed{n}\slashed{\epsilon}^*}{4}
\right)^{\beta\alpha},
\nonumber \\
B_{\rho\textrm{(d),light-cone}}(\ell^-)
&=&
\int d^{D-2}\ell_{\perp}
\partial_{\ell_{\perp}\rho}
\left[
\delta^{D-2}({\ell}_{\perp})
\right]  
(T^e)^{bc}(T^e)^{da}
\left\{
\left(
g_s \gamma^\perp_\mu
\right)
P_n
\Pi_{^3S_1^{[1]}}^{cd}
(ig_s\gamma_\nu)
\frac{i}
{2\slashed{p}-\slashed{\ell}-m+i\varepsilon}
P_{\bar{n}}
\right\}^{\beta\alpha}
\nonumber \\
&&
\times
\frac{i}{\left(p-\ell\right)^2+i\varepsilon}
\left[
-g^{\mu\nu}
+\frac{\left({p}-\ell\right)^\mu \bar{n}^\nu
+\bar{n}^\mu \left({p}-\ell\right)^\nu}
{\left({p}-\ell\right)\cdot \bar{n}}
\right]
\nonumber \\
&=&
\int d^{D-2}\ell_{\perp}
\partial_{\ell_{\perp}\rho}
\left[
\delta^{D-2}({\ell}_{\perp})
\right]  
\frac{ig_s^2C_F\delta^{ba}}{\sqrt{2N_c}mp^+}
\frac{1}{(-\ell^-+i\varepsilon)^2}
\left(
\frac{\slashed{\ell}_\perp
\slashed{n}
\slashed{\epsilon}^*}{2}
\right)^{\beta\alpha}
\nonumber \\
&=&
-
\frac{ig_s^2C_F\delta^{ba}}{\sqrt{2N_c}mp^+}
\frac{1}{(-\ell^-+i\varepsilon)^2}
\left(
\frac{\gamma_{\perp\rho}
\slashed{n}
\slashed{\epsilon}^*
}{2}
\right)^{\beta\alpha},
\nonumber \\
M_{\textrm{(d),light-cone}}(\ell^-)
&=&0,
\end{eqnarray}
where we have kept only the leading nonzero power of $\lambda$ in the
last line of each expression. Note that 
$M_{\textrm{(d),light-cone}}(\ell^-)=0$ because $V_{2m}^{(1)}=0$ 
in Fig.~\ref{fig:feyn-Ajpsi}.
By comparing the light-cone-gauge results in Eq.~(\ref{eq:light-cone-f})
with the Feynman-gauge results in Eq.~(\ref{eq:rad-Feyn-all}),
we conclude that all three radiative jet functions 
in Eq.~(\ref{def:A-3S1-BCDF}) are not gauge invariant. The
difference between the light-cone-gauge and the Feynman-gauge
calculations is
\begin{eqnarray}
\label{eq:lc-feynman-diff}%
&&[A_{\textrm{(d),light-cone}}(\ell^-)+
(-\ell_{\perp}^\rho) B_{\rho\textrm{(d),light-cone}}(\ell^-)
+M_{\textrm{(d),light-cone}}(\ell^-)]\nonumber\\
&&-
[A_{\textrm{(c),Feynman}}(\ell^-)+
(-\ell_{\perp}^\rho) B_{\rho\textrm{(d),Feynman}}(\ell^-)
+M_{\textrm{(a),Feynman}}(\ell^-)]
\nonumber \\
&=&
-
\frac{ig_s^2C_F\delta^{ba}}{\sqrt{2N_c}mp^+}
\frac{1}{(-\ell^-+i\varepsilon)^2}
\left[
\left(-\slashed{\ell}-m\right)
\frac{
\slashed{n}
\slashed{\epsilon}^*
}{2}
\right]^{\beta\alpha},
\end{eqnarray}
where we have contracted $(-\ell_{\perp}^\rho)$ into $B_\rho$ since an
additional factor of $(-\ell_{\perp}^\rho)$ is present in the soft
function that is associated with $B_\rho$, relative to the soft
functions that are associated with $A$ and $M$.  In the last line of
Eq.~(\ref{eq:lc-feynman-diff}), we have inserted $-\ell^+
\frac{\slashed{n}}{2} \slashed{n}=0$ in order to obtain the factor
$-\slashed{\ell}-m$. As is expected from our Ward-identity analysis, the
gauge-variant terms are proportional to the inverse of the soft-quark
propagator.

\subsection{Covariant gauge}

We can also consider the radiative jet functions in a general covariant
gauge, in which the gluon polarization sum is given by 
\begin{equation}
-g^{\mu\nu}+\alpha\frac{k^\mu k^\nu}{k^2},
\label{eq:gen-cov}
\end{equation}
where $\alpha$ is the gauge parameter.  

In order $\alpha_s$ in a general covariant gauge, the sum over all
connections of the gluon to the quark line and the $W_n$ Wilson line (on
the right side of a radiative-jet diagrams) is gauge invariant,
independently of the rest of the diagram.  Hence, the term in the
polarization sum that is proportional to $\alpha$ vanishes in the sum
over all connections of the gluon to the quark line and the $W_n$ Wilson
line, and one obtains the Feynman-gauge result.

However, in order $\alpha_s^2$ the situation is more complicated.  We
have checked that, in the Abelian case, there is a remnant from the
terms that are proportional to $\alpha$ in the polarization sum that is
nonzero at the integrand level.  This suggests that a gauge dependence
exists in general covariant gauges in order $\alpha_s^2$.

\section{Gauge invariant SCET Lagrangians \label{sec:gauge-inv-lagrangian}}
In this section, we modify the BCDF Lagrangians to obtain
    gauge-invariant Lagrangians that describe the coupling of a soft
quark to a hard-collinear quark.  In order to account for the
gauge-violating contribution in Eq.~(\ref{eq:gauge-violating}), we
introduce the following subtraction Lagrangian:
\begin{eqnarray}
\label{eq:del-L}
\Delta{\cal L}_{2}(x)
&=&
\left[
\bar{q}_s
\left(
-i\overleftarrow{\slashed{D}}_{s}
-m
\right)
\right]
(x^+)
(W_n^\dagger
\xi_n)(x)
\nonumber \\
&=&
\left[
\bar{q}_s
\left(
\frac{\slashed{\bar{n}}}{2} in\cdot D_s
-
i\overleftarrow{\slashed{D}}_{{s}\perp}
-
m
\right)
\right]
(x^+)
(W_n^\dagger
\xi_n)(x),
\end{eqnarray}
where, in the second line, we have used $\slashed{n} \xi_n = \slashed{n}
P_n \psi_n=0$, and performed an integration by parts for the term that is proportional to $in\cdot
D_s$. This integration by parts is valid because the minus
component of the soft momentum flows into the hard-collinear parts of
the Lagrangian.  We remind the reader that the position arguments of the
soft fields have been multipole expanded (depend only on the plus
component of the coordinate), but only after the derivatives acting on
the soft fields have been taken.

The modified Lagrangian that describes the coupling of a soft quark to
    a hard-collinear quark in order $\lambda^2$ is
\begin{eqnarray}
\label{eq:new-before-BPS}
{\cal L}_{2}^{\rm mod}
&=&
{\cal L}_{2a}^{\rm BCDF}
+
{\cal L}_{2b}^{\rm BCDF}
+
{\cal L}_{2m}^{\rm BCDF}
-
\Delta{\cal L}_{2}
\nonumber \\
&=&
\bar{q}_s
W_n^\dagger
in\cdot D
\frac{\slashed{\bar{n}}}{2}
\xi_n
-
\bar{q}_s
in\cdot D_{s}
W_n^\dagger
\frac{\slashed{\bar{n}}}{2}
\xi_n
+
\bar{q}_sW_n^\dagger
i\slashed{D}_{n\perp}
\frac{1}{i\bar{n}\cdot D_n} i\slashed{D}_{n\perp}
\frac{\slashed{\bar{n}}}{2}
\xi_n
\nonumber \\
&&
+
\bar{q}_s (-i\overleftarrow{D}_{s\perp})\cdot ix_{\perp} 
W_n^\dagger i\slashed{D}_{n\perp} \xi_n
-
\bar{q}_s
(-i\overleftarrow{\slashed{D}}_{s\perp})
W_n^\dagger
\xi_n,
\end{eqnarray}
where we omit the position arguments for simplicity.  Note that the
quark-mass-dependent Lagrangian ${\cal L}_{2m}^{\rm BCDF}$ is canceled
by the mass term of $\Delta \mathcal{L}_{2}$ in Eq.~(\ref{eq:del-L}),
and so the $O(\lambda^2)$ interactions between a soft quark and a
    hard-collinear quark are independent of the quark mass.  

\begin{figure}
\centering
\includegraphics[width=0.7\columnwidth]{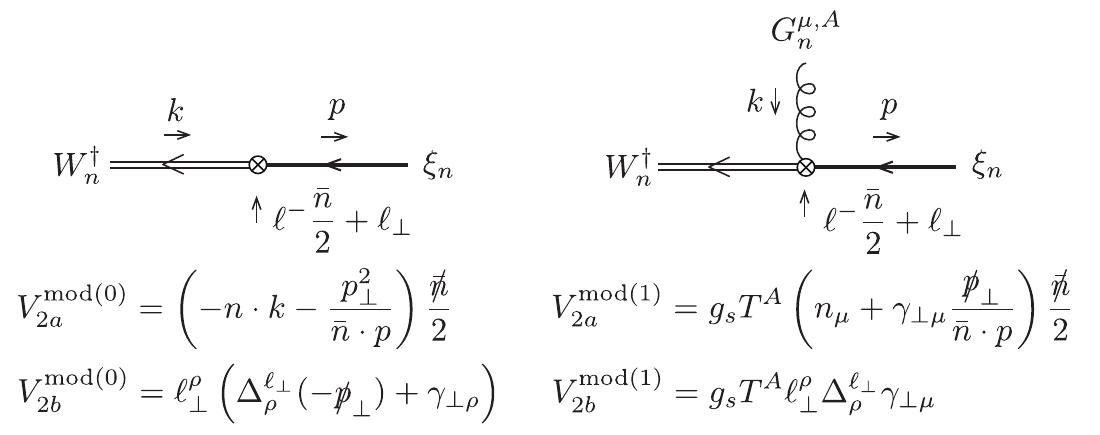}
\caption{\label{fig:feyn-Ajpsi3} \setstretch{1.5}The Feynman rules for the crossed
  circles in Fig.~\ref{fig:A1} for the Lagrangians in
  Eq.~(\ref{eq:new-lag-after-bps}).  $V_{2a}^\textrm{mod}$ and
  $V_{2b}^\textrm{mod}$ represent the crossed-circle contributions
      that arise from ${\cal L}_{2a}^{\rm mod, BPS}$ and
      ${\cal L}_{2b}^{\rm mod, BPS}$, respectively.  The
      superscripts $(0)$ and $(1)$ denote the order-$g_s^0$ and
      order-$g_s^1$ contributions of the crossed circle, respectively. }
\end{figure}

At this stage, longitudinally polarized soft gluons can still interact
    with the hard-collinear subdiagram.  As we have mentioned
    previously, such interactions can be factored into Wilson lines by
    making use of the BPS field redefinitions \cite{Bauer:2001yt}, which
    are shown in Eq.~(\ref{eq:BPS-redef}).  Using the BPS field
    redefinitions and the identities
\begin{subequations}%
\begin{eqnarray}
S_n^\dagger in\cdot D_{s} S_n&=&in\cdot \partial, 
\\
S_n^\dagger in\cdot D' S_n
&=& 
S_n^\dagger n\cdot 
\left(S_ng_sG_nS_n^\dagger + iD_s \right) 
S_n
=
in \cdot D_n, 
\end{eqnarray}
\end{subequations}%
we rewrite Eq.~(\ref{eq:new-before-BPS}) as
follows:
\begin{eqnarray}
\label{eq:position-PBS-final}
{\cal L}_{2}^{\rm mod, BPS}
&=&
{\cal L}_{2a}^{\rm mod, BPS}
+
{\cal L}_{2b}^{\rm mod, BPS},
\end{eqnarray}
where
\begin{subequations}%
\label{eq:new-lag-after-bps}%
\begin{eqnarray}
\label{eq:new-lag-after-bps-a}%
{\cal L}_{2a}^{\rm mod, BPS}
&=&
\bar{q}_s
S_n
\left[
W_n^\dagger
\left(
-
in\cdot \overleftarrow{D}_n  \right)
+
W_n^\dagger
i\slashed{D}_{n\perp}
\frac{1}{i\bar{n}\cdot D_n} i\slashed{D}_{n\perp}
\right]
\frac{\slashed{\bar{n}}}{2}
\xi_n,
\\
\label{eq:new-lag-after-bps-b}%
{\cal L}_{2b}^{\rm mod, BPS}
&=&
\bar{q}_s (-i\overleftarrow{D}_{s\perp}^\rho)
S_n 
\left(
ix_{\perp\rho} W_n^\dagger i\slashed{D}_{n\perp} \xi_n
-
\gamma_{\perp\rho}
W_n^\dagger
\xi_n
\right).
\end{eqnarray}
\end{subequations}
Here, in the first term of ${\cal L}_{2a}^{\rm mod, BPS}$, the
covariant derivative $(-in\cdot \overleftarrow{D}_n)$ should be
understood as acting only on the Wilson line $W_n^\dagger$.  The Feynman
rules for these modified Lagrangians are given in
Fig.~\ref{fig:feyn-Ajpsi3}.  Again, we note that, owing to the use of
the multipole expansion, the soft transverse momentum $\ell_\perp$
should be set to zero, except in the terms involving the operator
$\Delta_{\rho}^{\ell_\perp}$ [Eq.~(\ref{def:del-ellperp-op})].  For
those terms, $\ell_\perp$ is set to zero after differentiation with
respect to $\ell_\perp$.
    
\subsection{Gauge invariance and Ward identities of the modified Lagrangians}

Let us repeat the gauge-invariance analysis of
Sec.~\ref{sec:ward-original} for the modified $O(\lambda^2)$ SCET
Lagrangians in Eq.~(\ref{eq:new-lag-after-bps}). The amplitudes for
${\cal L}_{2a}^{\rm mod, BPS}$ and ${\cal L}_{2b}^{\rm mod,
  BPS}$, which correspond to the diagram in Fig.~\ref{fig:A1}, are
\begin{subequations}%
\begin{eqnarray}
\label{eq:ABBm-new}
A_{2a}^\textrm{mod}
&=&
\epsilon^\mu(k) \left[
\frac{(-ig_s)(i) \bar{n}_\mu}{\bar{n}\cdot k
+i\varepsilon}  
\left(
-n\cdot k-\frac{{p}_\perp^2}{\bar{n}\cdot p}
\right)
+ g_s
\left(
n_\mu
+
\gamma_{\perp\mu}
\frac{\slashed{p}_\perp}{\bar{n}\cdot p}
\right)
\right]\frac{\slashed{\bar{n}}}{2} P_n
\ldots,
\\
A_{2b}^\textrm{mod}
&=&
\epsilon^\mu(k) 
\left[\frac{(-ig_s)(i) \bar{n}_\mu}{\bar{n}\cdot k
+i\varepsilon}
\ell_{\perp}^\rho
\left(
\Delta^{\ell_\perp}_{\rho}
(-\slashed{p}_\perp)
+\gamma_{\perp\rho}
\right)
+
g_s
\ell_{\perp}^\rho
\Delta^{\ell_\perp}_{\rho}
\gamma_{\perp\mu}
\right]
P_n\ldots,
\end{eqnarray}
\end{subequations}%
respectively. We obtain the Ward identities for these amplitudes by
replacing the polarization vector $\epsilon^{\mu}(k)$ with the
factor $k^\mu$ from the gauge transformation in
Eq.~(\ref{eq:gauge-transform}).  The results are as follows:
\begin{subequations}%
\label{eq:ward-ident-new}%
\begin{eqnarray}
A_{2a}^\textrm{mod,gauge}
&=&
g_s
\left[
\left(
-n\cdot k-\frac{{p}_\perp^2}{\bar{n}\cdot p}
\right)
+
\left(
n\cdot k
+
\frac{\slashed{k}_\perp\slashed{p}_\perp}{\bar{n}\cdot p}
\right)
\right]\frac{\slashed{\bar{n}}}{2} P_n
\ldots
\\
\label{eq:ward-ident-new-b}%
&=&
0,
\nonumber \\
A_{2b}^\textrm{mod,gauge}
&=&
g_s
\left[
\ell_{\perp}^\rho
\left(
\Delta^{\ell_\perp}_{\rho}
(-\slashed{p}_\perp)
+\gamma_{\perp\rho}
\right)
+
\ell_{\perp}^\rho
\Delta^{\ell_\perp}_{\rho}
\slashed{k}_\perp
\right]
P_n\ldots
\nonumber \\
&=&
0,
\end{eqnarray}
\end{subequations}%
where we have used $k_{\perp} = p_{\perp}$ for
$A_{2a}^\textrm{mod,gauge}$, owing to the multipole expansion of the
soft-quark field, and $k_{\perp} = p_{\perp}-\ell_\perp$ for
$A_{2b}^\textrm{mod,gauge}$, owing to the insertion of $\ell_\perp$ that
follows from the identity in Eq.~(\ref{eq:identity-ix}).  Note that
    the contributions in which $\Delta_\rho^{\ell_\perp}$ acts on the
    ellipsis (the remainder of the diagram) cancel between the first and
    third terms after the first equality in
    Eq.~(\ref{eq:ward-ident-new-b}).  We conclude that the modified
  Lagrangians in Eq.~(\ref{eq:new-lag-after-bps}) are separately gauge
  invariant.

\subsection{Missing diagrams for the modified Lagrangians \label{sec:missing-new}}
Now let us consider the contributions of the modified Lagrangians to the
missing diagram in Fig.~\ref{fig:A2}.  The amplitudes for the diagram in
Fig.~\ref{fig:A2} that correspond to the crossed vertices from ${\cal
  L}_{2a}^{\rm mod, BPS}$ and ${\cal L}_{2b}^{\rm mod, BPS}$ are
\begin{subequations}%
\label{eq:ABBm2-new-soft}%
\begin{eqnarray}
A_{2a,\textrm{miss}}^\textrm{mod}
&=&0,
\\
A_{2b,\textrm{miss}}^\textrm{mod}&=&
\epsilon^\mu(k) 
\ell_\perp^\rho
\left(
\Delta_\rho^{\ell_\perp}
(-\slashed{\ell}_\perp)
+\gamma_{\perp\rho}
\right)
P_n
\frac{i}{-\slashed{\ell}-m+i\varepsilon}
(ig_s) \gamma_\mu 
\ldots
\nonumber \\
&=&0,
\end{eqnarray}
\end{subequations}%
respectively. Here, for $A_{2a,\textrm{miss}}^\textrm{mod}$, we have performed
the multipole expansion to set $\ell_\perp$ to $0$, and 
for $A_{2b,\textrm{miss}}^\textrm{mod}$, we have used
\begin{equation}
\ell_\perp^\rho
\left(\Delta_\rho^{\ell_\perp}
(-\slashed{\ell}_\perp)
+\gamma_{\perp\rho}
\right)
=
\left(
\ell_\perp^\rho 
\Delta_\rho^{\ell_\perp}
(-\slashed{\ell}_\perp)
+\slashed{\ell}_\perp
\right)
=0.
\end{equation}
We see that the modified Lagrangians in
Eq.~(\ref{eq:new-lag-after-bps}) give vanishing contributions in order
$g_s$ to the missing diagram in Fig.~\ref{fig:A2} and are, therefore,
gauge invariant to this order.

\section{Gauge invariance to all orders in 
$g_s$\label{sec:all-orders-gauge-inv}}

\begin{figure}
\centering
\includegraphics[scale=0.9]{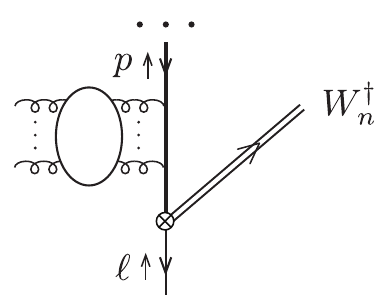}
\caption{\label{fig:A3} 
\setstretch{1.5}The general form of the missing diagrams at all orders in $g_s$.
The blob includes gluon self interactions and ghost loops.}
\end{figure}

We now argue that the gauge invariances of order-$\lambda^1$ Lagrangian
in Eq.~(\ref{eq:collinear-soft-all-1}) and the modified order-$\lambda^2$
Lagrangians in Eq.~(\ref{eq:new-lag-after-bps}) hold to all orders in
$g_s$.  First, we note that the all-orders missing diagrams have exactly
one soft-quark propagator and that all collinear gluon attachments are to the
collinear-subdiagram side of the soft-quark propagator. These diagrams
are of the form that is shown in Fig.~\ref{fig:A3}. A crucial feature of
these diagrams is that the crossed-vertex factors are independent of the
hard-collinear gluon attachments. That is, they are equal to the
lowest-order crossed-vertex factors. Furthermore, we have seen in
Eqs.~(\ref{eq:ABBm2-a}) and (\ref{eq:ABBm2-new-soft}) that each
crossed-vertex factor gives a vanishing result when no hard-collinear
momentum flows through the crossed vertex. Therefore, the missing
diagrams give vanishing contributions, ensuring the gauge invariance of
the Lagrangians in Eqs.~(\ref{eq:collinear-soft-all-1}) and
(\ref{eq:new-lag-after-bps}).

\section{Radiative jet function in order $\alpha_s$
from the modified Lagrangians \label{sec:radiative-new}} In this section,
        we consider radiative jet functions that arise from
        the gauge-invariant Lagrangians in
        Eq.~(\ref{eq:new-lag-after-bps}).  Using the two
            $O(\lambda^2)$ gauge-invariant Lagrangians, we
        construct the following radiative jet functions:
\begin{eqnarray}
\label{def:A-3S1-new}
A^\textrm{mod}(\ell^-)
&=&
\int d^D x\,
e^{-i\frac{\ell^- x^+}{2}}
\langle {\cal Q}{\cal\bar{Q}}(^3S_1^{[1]},p,p) | 
\nonumber \\
&&
\times
T\,
\left\{
\left[
W_n^\dagger
\left(
-
in\cdot \overleftarrow{D}_n  \right)
+
W_n^\dagger
i\slashed{D}_{n\perp}
\frac{1}{i\bar{n}\cdot D_n} i\slashed{D}_{n\perp}
\right]
\frac{\slashed{\bar{n}}}{2}
\xi_n
\right\}^{\beta,b}(x)
(\bar{\xi}_nW_n)^{\alpha,a}(0)
|0\rangle,
\nonumber \\
B_{1\rho}^\textrm{mod}(\ell^-)
&=&
\int d^{D-2}\ell_{\perp}
\frac{\partial}{\partial\ell_{\perp}^\rho}
\left[
\delta^{D-2}({\ell}_{\perp})
\right]  
\nonumber \\
&&
\times
\int d^D x\,
e^{-i\frac{\ell^- x^+}{2}}
e^{-i\ell_\perp\cdot x_\perp}
\langle {\cal Q}{\cal\bar{Q}}(^3S_1^{[1]},p,p) | 
T\,
\left(W_n^\dagger i\slashed{D}_{n\perp} \xi_n\right)^{\beta,b}(x)
(\bar{\xi}_nW_n)^{\alpha,a}(0)
|0\rangle,
\nonumber \\
B^{\textrm{mod}}_{2\rho}(\ell^-) 
&=&
\int d^D x\,
e^{-i\frac{\ell^- x^+}{2}}
\langle {\cal Q}{\cal\bar{Q}}(^3S_1^{[1]},p,p) |  
T
\left(
-
\gamma_{\perp\rho}
W_n^\dagger
\xi_n
\right)^{\beta,b}(x)
(\bar{\xi}_nW_n)^{\alpha, a}(0)
|0\rangle.
\end{eqnarray}
Here, we have split the contribution from ${\cal L}_{2b}^{\rm mod, BPS}$
in Eq.~(\ref{eq:new-lag-after-bps}) into the contributions
$B_{1\rho}^\textrm{mod}$ and $B_{2\rho}^\textrm{mod}$ because these
contributions contain different prescriptions for the multipole
expansion of the soft momentum $\ell$. In accordance with our
Ward-identity result in Eq.~(\ref{eq:ward-ident-new-b}), we will find
that only the sum of $B_{1\rho}^\textrm{mod}$ and
$B_{2\rho}^\textrm{mod}$ is gauge invariant.

In the Feynman gauge, we find that the diagrams of
Figs.~\ref{fig:AA-LO}(a) and (c) contribute to the $A^\textrm{mod}$
radiative jet function. (When we use the modified Feynman rules in
Fig.~\ref{fig:feyn-Ajpsi3}, diagram (a) is no longer power suppressed.)
We also find that the sole contribution to the $B_{1\rho}^\textrm{mod}$
radiative jet function arises from the diagram of
Fig.~\ref{fig:AA-LO}(d), and the sole contribution to the
$B_{2\rho}^\textrm{mod}$ radiative jet function arises from the
diagram of Fig.~\ref{fig:AA-LO}(a).  (The reasoning is the same as for
the $B$ and $M$ radiative jet functions in Sec.~\ref{sec:feyn-BCDF}.) By
making use of the Feynman rules in Fig.~\ref{fig:feyn-Ajpsi3}, we obtain
\begin{eqnarray}
\label{eq:rad-Feyn-all-new}
A_{\textrm{(a),Feynman}}^\textrm{mod}(\ell^-)
&=&
(T^e)^{bc}(T^e)^{da}
\left\{
\left[
-n\cdot (p-\ell)\frac{\slashed{\bar{n}}}{2}
\right]
P_n
\Pi_{^3S_1^{[1]}}^{cd}
(ig_s\gamma_\mu)
\frac{i}{2\slashed{p}-\slashed{\ell}-m+i\varepsilon}
P_{\bar{n}}
\right\}^{\beta\alpha}
\nonumber \\
&&
\times
\frac{i(-ig_s \bar{n}^\mu)}
{\bar{n}\cdot \left(p-\ell\right)+i\varepsilon}
\frac{-i}{\left(p-\ell\right)^2+i\varepsilon}
\nonumber \\
&=&
-
\frac{ig_s^2C_F\delta^{ba}}{\sqrt{2N_c}mp^+}
\frac{1}
{(-\ell^-+i\varepsilon)}
\left(
\frac{\slashed{\bar{n}}\slashed{n}\slashed{\epsilon}^*}{4}
\right)^{\beta\alpha},
\nonumber \\
A_{\textrm{(c),Feynman}}^\textrm{mod}(\ell^-)
&=&
A_{\textrm{(c),Feynman}}(\ell^-)
\nonumber \\
&=&
-
\frac{ig_s^2C_F\delta^{ba}}{\sqrt{2N_c}mp^+}
\frac{1}
{(-\ell^- +i\varepsilon)}
\left(
\frac{\slashed{\bar{n}}\slashed{n}\slashed{\epsilon}^*}{4}
\right)^{\beta\alpha},
\nonumber \\
B_{1\rho,\textrm{(d),Feynman}}^\textrm{mod}(\ell^-)
&=&
B_{\rho\textrm{(d),Feynman}}(\ell^-)
\nonumber \\
&=&0,
\nonumber \\
B_{2\rho,\textrm{(a),Feynman}}^\textrm{mod}(\ell^-)
&=&
(T^e)^{bc}(T^e)^{da}
\left\{
\left(
-\gamma_{\perp\rho}
\right)
P_n
\Pi_{^3S_1^{[1]}}^{cd}
(ig_s\gamma_\mu)
\frac{i}
{2\slashed{p}-\slashed{\ell}-m+i\varepsilon}
P_{\bar{n}}
\right\}^{\beta\alpha}
\nonumber \\
&&
\times
\frac{i(-ig_s \bar{n}_\nu)}
{\bar{n}\cdot \left(p-\ell\right)+i\varepsilon}
\frac{-ig^{\mu\nu}}{\left(p-\ell\right)^2+i\varepsilon}
\nonumber \\
&=&
-
\frac{ig_s^2C_F\delta^{ba}}{\sqrt{2N_c}mp^+}
\frac{1}{(-\ell^-+i\varepsilon)^2}
\left(
\frac{\gamma_{\perp\rho}\slashed{n}\slashed{\epsilon}^*}{2}\right)^{\beta\alpha}.
\end{eqnarray}
We note that
    $A_{\textrm{(c),Feynman}}^\textrm{mod}(\ell^-)=A_{\textrm{(c),Feynman}}(\ell^-)$
    and
    $B_{1\rho,\textrm{(d),Feynman}}^\textrm{mod}(\ell^-)=B_{1\rho,\textrm{(d),Feynman}}(\ell^-)$,
    which follows from $V_{2a}^\textrm{mod(1)}=V_{2a}^{(1)}$ and
    $V_{2b}^\textrm{mod(1)} = V_{2b}^{(1)}$
    (Figs.~\ref{fig:feyn-Ajpsi}, \ref{fig:feyn-Ajpsi2},
    and \ref{fig:feyn-Ajpsi3}).

In the light-cone gauge, only the diagram of
Fig.~\ref{fig:AA-LO}(d) contributes to the radiative jet
function. We find that
\begin{eqnarray}
A_{\textrm{(d),light-cone}}^\textrm{mod}(\ell^-)
&=&
A_{\textrm{(d),light-cone}}(\ell^-)
\nonumber \\
&=&
-
\frac{2ig_s^2C_F\delta^{ba}}{\sqrt{2N_c}mp^+}
\frac{1}{(-\ell^-+i\varepsilon)}
\left(
\frac{\slashed{\bar{n}}\slashed{n}\slashed{\epsilon}^*}{4}
\right)^{\beta\alpha},
\nonumber \\
B_{1\rho,\textrm{(d),light-cone}}^\textrm{mod}(\ell^-)
&=&
B_{\rho\textrm{(d),light-cone}}(\ell^-)
\nonumber \\
&=&
-
\frac{ig_s^2C_F\delta^{ba}}{\sqrt{2N_c}mp^+}
\frac{1}{(-\ell^-+i\varepsilon)^2}
\left(
\frac{\gamma_{\perp\rho}
\slashed{n}
\slashed{\epsilon}^*
}{2}
\right)^{\delta\alpha},
\nonumber \\
B_{2\rho,\textrm{(d),light-cone}}^\textrm{mod}(\ell^-)
&=&0,
\end{eqnarray}
where we have used the results in Eq.~(\ref{eq:light-cone-f}), since
$V_{2a}^\textrm{mod(1)}=V_{2a}^{(1)}$ and $V_{2b}^\textrm{mod(1)} =
V_{2b}^{(1)}$ (Figs.~\ref{fig:feyn-Ajpsi}, \ref{fig:feyn-Ajpsi2}, and
    \ref{fig:feyn-Ajpsi3}).  Note that
$B_{2\rho,\textrm{(d),light-cone}}^\textrm{mod}(\ell^-)=0$ because, as can be seen
    from Fig.~\ref{fig:feyn-Ajpsi3}, the relevant order-$g_s$ vertex
vanishes.

We conclude that the radiative jet functions are identical in the
    Feynman gauge and the light-cone gauge:
\begin{eqnarray}
A_{\textrm{(a),Feynman}}^\textrm{mod}(\ell^-)
+
A_{\textrm{(c),Feynman}}^\textrm{mod}(\ell^-)
&=&
A_{\textrm{(d),light-cone}}^\textrm{mod}(\ell^-),
\nonumber \\
B_{1\rho,\textrm{(d),Feynman}}^\textrm{mod}(\ell^-)
+
B_{2\rho,\textrm{(a),Feynman}}^\textrm{mod}(\ell^-)
&=&
B_{1\rho,\textrm{(d),light-cone}}^\textrm{mod}(\ell^-)
+
B_{2\rho,\textrm{(d),light-cone}}^\textrm{mod}(\ell^-).
\end{eqnarray}
As is expected from our Ward-identity analysis
    [Eq.~(\ref{eq:ward-ident-new})], $A^\textrm{mod}$ is separately
gauge invariant, while the sum of $B_{1\rho}^\textrm{mod}$ and
$B_{2\rho}^\textrm{mod}$ is gauge invariant.

\section{Gauge invariance in the label-momentum formulation of SCET
\label{sec:label-momentum}}

Finally, we mention that certain versions of the label-momentum
formulation of SCET \cite{Pirjol:2002km,Bauer:2003mga} also contain
expressions for the interactions between a soft-quark and a
hard-collinear quark that are constructed from ostensibly
gauge-invariant operators.  The final expressions for the
soft-quark-to-hard-collinear-quark Lagrangians are given in Eqs.~(33)
and (35) of Ref.~\cite{Pirjol:2002km} and Eq.~(27) of
Ref.~\cite{Bauer:2003mga}. They are derived by making use of the
soft-quark equation of motion, which, as we have seen, is a crucial
ingredient in deriving a gauge-invariant Lagrangian that describes
the transitions of a soft quark to a hard-collinear quark.

The soft-quark-to-hard-collinear-quark Lagrangians in Eqs.~(33) and (35)
of Ref.~\cite{Pirjol:2002km} are proportional to the quantities
$\slashed{M}$ and $\slashed{B}_\perp^c$, and the
soft-quark-to-hard-collinear-quark Lagrangians in Eq.~(27) of
Ref.~\cite{Bauer:2003mga} are proportional to the quantities and $n\cdot
M$ and $\slashed{B}_\perp^c$. Since these quantities are commutators of
covariant derivatives, the corresponding Lagrangians vanish when the
gauge fields are set to zero.  That is, 
these Lagrangians must produce at least one gluon emission if they
are to give nonvanishing contributions to amplitudes.
Consequently, the missing diagrams (Fig.~\ref{fig:A3}) receive vanishing
contributions from these Lagrangians, and these Lagrangians evade the
gauge-invariance issue that we have identified in this paper.  That is,
the corresponding operators are truly gauge invariant.

As a further check, we have used the Lagrangians in Eqs.~(33) and (35) of
Ref.~\cite{Pirjol:2002km}, to compute radiative jet functions in order
$\alpha_s$ and found agreement with our results for the radiative jet
functions in Sec.~\ref{sec:radiative-new}.  Here, it was necessary to
sum the contributions from the order-$\lambda$ operator in Eq.~(33) and
the $\slashed{M}_\perp$ order-$\lambda^2$ operator in Eq.~(35) in
order to obtain gauge-invariant results, in accordance with the
observation in Ref.~\cite{Bauer:2003mga} that collinear gauge
transformations of these Lagrangians mix operators that have different
scaling in $\lambda$.

We have also used the Lagrangians in Eq.~(27) of
Ref.~\cite{Bauer:2003mga} to compute radiative jet functions in order
$\alpha_s$. Again, we have found agreement with our results for the
radiative jet functions in Sec.~\ref{sec:radiative-new}. In this
computation, we used the Feynman rules that are given in the erratum to
Ref.~\cite{Pirjol:2002km}.  In order to obtain all of the contributions
to the radiative jet functions of order $\lambda^2$, it was also necessary, in the
case of these Lagrangians, to consider the order-$\lambda$ correction to
the gluon propagator.\footnote{\setstretch{0.8}This contribution to the radiative jet
  functions corresponds to the contribution of the $\slashed{M}_\perp$
  term in Eq.~(35) of Ref.~\cite{Pirjol:2002km}. The $\slashed{M}_\perp$
  term is absent in the Lagrangian in Eq.~(27) of
  Ref.~\cite{Bauer:2003mga}.}  We derived the Feynman rule for this
correction to the gluon propagator by applying the field redefinitions
in Eq.~(14) of Ref.~\cite{Bauer:2003mga} to the gauge-fixed gauge-field
action.

\section{Discussion}
\label{sec:discussion}

In this paper we have pointed out that the Lagrangians in
    Refs.~\cite{Beneke:2002ph, Beneke:2002ni} (BCDF) are not gauge
    invariant in the hard-collinear sector.  This is surprising because
    these Lagrangians are constructed from operators that are ostensibly
    gauge invariant: hard-collinear fields are accompanied by
    Wilson-line factors, and all derivatives are covariant
    derivatives. The violations of gauge invariance are somewhat
    subtle. They arise because hard-collinear gauge transformations
    multiply the quark fields by a phase that, in momentum space, can
    shift the quark-field momentum from the hard-collinear region to the
    soft region, where the hard-collinear quark field, by definition,
    has no support. This phenomenon is manifested in perturbation theory
    in the hard-collinear sector through the absence of certain diagrams
    that would be present in full QCD.  One consequence of the absence
    of these diagrams is that, if one uses the BCDF Lagrangians directly
    to construct radiative jet functions, then the radiative jet
    functions are not gauge invariant by themselves.
    
We have demonstrated the violations of gauge invariance by examining
    the Ward identities for the BCDF Lagrangians and also by computing
    the radiative jet functions that follow directly from the BCDF
    Lagrangians at the leading order in $g_s$ in the Feynman gauge and
    in the light-cone gauge.  These analyses show that the violations of
    gauge invariance are, at the leading nontrivial
    order in $g_s$, proportional to inverse of the soft-quark
    propagator.

Motivated by the Ward-identity and radiative-jet-function analyses,
we have modified the BCDF Lagrangians by adding terms that are
proportional to the soft-quark equation of motion.  Then, after making
use of the BPS field redefinition to factor minus-polarized soft gluons
from the hard-collinear subdiagram, we have arrived at gauge-invariant
Lagrangians, through order $\lambda^2$, that describe the couplings of a
soft quark to a hard-collinear quark. The fact that the violations of
gauge invariance can be removed through the use of a field redefinition
that is proportional to the soft-quark equation of motion implies that
$S$-matrix elements of the original BCDF Lagrangians are gauge
invariant.

The modified gauge-invariant Lagrangians that we have derived are
    somewhat more general than the BCDF Lagrangians, in that we have
    considered the case of a nonzero quark mass. We have demonstrated
    the gauge invariance of the modified Lagrangians through examination
    of Ward identities in order $g_s$ and through calculations of
    radiative jet functions in order $\alpha_s$ in the Feynman gauge and
    the light-cone gauge. We have also given a Ward-identity argument
    to show that the gauge invariance holds to all orders in $g_s$.

In Refs.~\cite{Liu:2019oav,Liu:2020ydl}, the order-$\lambda^1$
Lagrangian in Eq.~(\ref{eq:collinear-soft-all-1}) was used to construct
a radiative jet function that involves a single-photon external state,
and that radiative jet function was computed at one and two loops in
perturbation theory in the light-cone gauge.\footnote{\setstretch{0.8}When the
    transition of soft quark to a hard-collinear quark involves the
    emission of a single external photon, the corresponding radiative
jet function of leading order in $\lambda$ derives from the
order-$\lambda^1$ Lagrangian in Eq.~(\ref{eq:collinear-soft-all-1}).  In
this case, the order-$\lambda^2$ Lagrangians in
Eq.~(\ref{eq:new-lag-after-bps}) yield radiative jet functions that are
suppressed by at least one power of $\lambda$.}  As we have seen in
Sec.~\ref{sec:all-orders-gauge-inv}, the Lagrangian in
Eq.~(\ref{eq:collinear-soft-all-1}) is gauge invariant to all orders in
perturbation theory. Consequently, the calculations in
Refs.~\cite{Liu:2019oav,Liu:2020ydl} are gauge invariant.

We have also examined the label-momentum formulation of SCET in the
incarnations that are given in Refs.~\cite{Pirjol:2002km,Bauer:2003mga}.
The Lagrangians in these papers that describe the interactions between a
soft quark and a hard-collinear quark contain commutators of covariant
derivatives.  Consequently, these Lagrangians must produce at least
    one gluon emission if they are to give nonvanishing contributions to
    amplitudes.  It follows that the missing diagrams that correspond to
    these Lagrangians vanish and that the formulations of SCET in
    Refs.~\cite{Pirjol:2002km,Bauer:2003mga} evade the gauge-invariance
    problem that we have identified in this paper.

\noindent {\bf Note added}

After the present paper appeared on the arXiv, a paper \cite{Boer:2023yde}
    was submitted to the arXiv that addresses the same gauge invariance
    issue that we address.  That paper presents the gauge-invariance
    issue from an alternative point of view in which the hard-collinear
    quark fields are not constrained to carry a hard-collinear
    momentum.

\begin{acknowledgments}
We wish to thank Martin Beneke, Philipp B\"oer, Hee Sok Chung, Patrick
Hager, Jungil Lee, Iain Stewart, and Yunlu Wang for helpful discussions.
D.K.\ thanks the Erwin-Schrödinger International Institute for Mathematics and Physics at the University of Vienna for partial support during the Program ``Quantum Field Theory at the Frontiers of the Strong Interactions", July 31 - September 1, 2023.
X.-P.\ W.\ would like to express special thanks to the Mainz Institute
for Theoretical Physics (MITP) of the Cluster of Excellence PRISMA+
(Project ID 39083149) for its hospitality and support.  The work of
G.T.B.\ is supported by the U.S.\ Department of Energy, Division of High
Energy Physics, under Contract No.\ DE-AC02-06CH11357.  The work of
D.K.\ and J.-H.E.\ is supported by the National Key Research and
Development Program of China under Contract No.~2020YFA0406301 and by
the National Natural Science Foundation of China (NSFC) through Grant
Nos.~12150610461, 11875112, and 12105051.  The work of X.-P.\ W.\ is
supported by the DFG (Deutsche Forschungsgemeinschaft, German Research
Foundation) Grant No.\ BR 4058/2-2 and by the the DFG cluster of
excellence “ORIGINS” under Germany’s Excellence Strategy - EXC-2094 -
390783311.  The submitted manuscript has been created in part by
UChicago Argonne, LLC, Operator of Argonne National Laboratory. Argonne,
a U.S.\ Department of Energy Office of Science laboratory, is operated
under Contract No.\ DE-AC02y-06CH11357. The U.S.\ Government retains for
itself, and others acting on its behalf, a paid-up nonexclusive,
irrevocable worldwide license in said article to reproduce, prepare
derivative works, distribute copies to the public, and perform publicly
and display publicly, by or on behalf of the Government.

All authors contributed equally to this work.
\end{acknowledgments}

\appendix
\section{Derivation of the mass-dependent SCET Lagrangians}
\label{app:mass-lag}
In this Appendix, we discuss the derivation of the mass-dependent
Lagrangian ${\cal L}_{2m}^{\rm BCDF}$ in
Eq.~(\ref{eq:collinear-soft-all}).  We follow the general outline of
    the analysis in Ref.~\cite{Beneke:2002ph}, but we start with the
full QCD Lagrangian with a quark mass, namely,
\begin{equation}
\mathcal{L} = \bar{\psi}(i\slashed{D}-m)\psi.
\end{equation}
We note that $m$ scales as $\lambda^2$.  We decompose this
Lagrangian in terms of the SCET fields, using $\psi =
\xi_n+\eta_n+q_s$. Then, we use the equation of motion for the small
component of the hard-collinear field $\eta_n$ to eliminate $\eta_n$
from the Lagrangian.  This leads to the following mass-dependent
Lagrangian, in addition to the terms in Eq.~(34) of
Ref.~\cite{Beneke:2002ph}:
\begin{eqnarray}
\label{eq:mass-dependent-lag}
\mathcal{L}_m
&=&
-m\bar{q}_sq_s
+
m\bar{\xi}_n \frac{1}{i\bar{n}\cdot D}
\frac{\slashed{\bar{n}}}{2}i\slashed{D}_\perp \xi_n
+
m\bar{\xi}_n i\slashed{D}_\perp
\frac{1}{i\bar{n}\cdot D}
\frac{\slashed{\bar{n}}}{2}\xi_n
\nonumber \\
&&
-
m^2
\bar\xi_n
\frac{1}{i\bar{n}\cdot D}
\frac{\slashed{\bar{n}}}{2}
\xi_n
+m
\bar\xi_n
\frac{1}{i\bar{n}\cdot D}
g_s 
\bar{n}\cdot G_n
q_s
+m
\bar q_s
g_s 
\bar{n}\cdot G_n
\frac{1}{i\bar{n}\cdot D}
\xi_n
\nonumber \\
&&
+
m
\bar\xi
\frac{1}{i\bar{n}\cdot D}
\frac{\slashed{\bar{n}}}{2}
g_s
\slashed{G}_{n\perp}
q_s
+m
\bar q_s
g_s 
\slashed{G}_{n\perp}
\frac{1}{i\bar{n}\cdot D}
\frac{\slashed{\bar{n}}}{2}
\xi,
\end{eqnarray}
where
\begin{eqnarray}
i\bar{n}\cdot D 
&=&
i\bar{n}\cdot \partial
+
g_s\bar{n}\cdot G_n
+
g_s\bar{n}\cdot G_s,
\nonumber \\
iD_\perp^\mu
&=&
iD_{n\perp}^\mu 
+g_s G_{s\perp}^\mu.
\end{eqnarray}
The first term of Eq.~(\ref{eq:mass-dependent-lag})
gives the order-$\lambda^0$ contribution to the action, 
and it yields the mass term of the leading-power soft-quark Lagrangian
\begin{equation}
\mathcal{L}_s^{(0)} = 
\bar{q}_s (i\slashed{D}_s -m)q_s.
\end{equation}
Note that the second and third terms on the right side of
    Eq.~(\ref{eq:mass-dependent-lag}) are of order $\lambda$ or smaller,
    while the fourth through sixth terms are of order $\lambda^2$, and
    the seventh and eighth terms are of order $\lambda^3$.

Next, let us perform the expansions in powers of $\lambda$ that are
required for homogeneous scaling for each term of the
Lagrangian. We can accomplish this by making use of the identity in
Eq.~(36) of Ref.~\cite{Beneke:2002ph},
\begin{eqnarray}
\label{eq:iden-phius}
\frac{1}{i\bar{n}\cdot D}
g_s \bar{n}\cdot G_n q_s
=
(1-WZ^\dagger)q_s
-\frac{1}{i\bar{n}\cdot D}
(1-WZ^\dagger)
i\bar{n}\cdot D_s q_s,
\end{eqnarray}
and the power expansions of the Wilson lines and the covariant derivatives,
\begin{eqnarray}
WZ^\dagger &=& W_n\left[1+O(\lambda^2)\right],
\nonumber \\
\frac{1}{i\bar{n}\cdot D}
&=&
\frac{1}{i\bar{n}\cdot D_n}
\left[
1
+O(\lambda^2)
\right],
\nonumber \\
iD_\perp^\mu
&=&
iD_{n\perp}^\mu 
\left[1
+O(\lambda)
\right], 
\end{eqnarray}
where we have given the relative orders in $\lambda$ of the corrections.
Then, through order $\lambda^2$, we obtain the following mass-dependent
Lagrangians:
\begin{eqnarray}
\mathcal{L}_m^{(0)}
&=&
-m\bar{q}_sq_s,
\nonumber \\
\mathcal{L}_m^{(1)}
&=&
m\bar{\xi}_n \frac{1}{i\bar{n}\cdot D_n}
\frac{\slashed{\bar{n}}}{2}i\slashed{D}_{n\perp} \xi_n
+
m\bar{\xi}_n i\slashed{D}_{n\perp}
\frac{1}{i\bar{n}\cdot D_n}
\frac{\slashed{\bar{n}}}{2}\xi_n,
\nonumber \\
\mathcal{L}_m^{(2)}
&=&
-
m^2
\bar\xi_n
\frac{1}{i\bar{n}\cdot D_n}
\frac{\slashed{\bar{n}}}{2}
\xi_n
-m
\bar{q}_s
W_n^{\dagger}
\xi_n
-m
\bar{\xi}_n W_n
q_{s}.
\end{eqnarray}
Here, in order to ensure homogeneous scaling in $\lambda$, one should
    multipole expand the position arguments of $q_s$ and $\bar{q}_s$ in
    $\mathcal{L}_m^{(2)}$ as follows:
\begin{equation}
q_s(x)
=q_s(x^+)
+[(x_\perp\cdot \partial_\perp) q_s](x^+)
+O(\lambda^2q_s).
\end{equation}
Note that the mass terms in $\mathcal{L}_m^{(1)}$ and the first
term of $\mathcal{L}_m^{(2)}$ would have given the mass term of the
leading-power SCET Lagrangian if we had used the power counting
$m\sim \lambda$, rather than $m\sim \lambda^2$, as would be appropriate
for collinear modes, rather than hard-collinear modes.  This would have
led to the following leading-power Lagrangian:
\begin{eqnarray}
\label{eq:lagrangian-with-mass}
\mathcal{L}^{(0)}_n\big|_{m\sim O(\lambda)}
&=&
\bar\xi_n
\left[
in\cdot D_n
+
\left(i\slashed{D}_{n\perp}
-m\right)
\frac{1}{i\bar{n}\cdot D_n}
\left(i\slashed{D}_{n\perp}
+m\right)
\right]
\frac{\slashed{\bar{n}}}{2}
\xi_n.
\end{eqnarray}

\section{Power suppressed diagrams of $A(\ell^-)$ in Feynman gauge}
\label{app:power-suppressed}
In this Appendix, we show that the diagrams of
    Figs.~\ref{fig:AA-LO}(a) and (d) lead to power-suppressed
    contributions in the Feynman gauge. 

The diagram of Fig.~\ref{fig:AA-LO}(a) gives
\begin{eqnarray}
\label{eq:diagram-a}%
&&
A_{\textrm{(a),Feynman}}(\ell^-)
\nonumber \\
&=&
(T^e)^{bc}
(T^e)^{da}
\left\{
\left(
-n\cdot p\frac{\slashed{\bar{n}}}{2}
\right)
P_n
\Pi_{^3S_1^{[1]}}^{cd}
(ig_s\gamma_\mu)
\frac{i}{2\slashed{p}-\slashed{\ell}-m+i\varepsilon}
P_{\bar{n}}
\right\}^{\beta\alpha}
\frac{i(-ig_s \bar{n}^\mu)}
{\bar{n}\cdot \left(p-\ell\right)+i\varepsilon}
\frac{-i}{\left(p-\ell\right)^2+i\varepsilon}
\nonumber \\
&=&
-\frac{ig_s^2C_F\delta^{ba}
}
{\sqrt{2N_c}mp^+}
\frac{1}{(-\ell^-+i\varepsilon)}
\left(
\frac{\slashed{\bar{n}}\slashed{n}\slashed{\epsilon}^*}{4}
\right)^{\beta\alpha}
\frac{p^-}{(-\ell^-+i\varepsilon)},
\end{eqnarray}
where we have kept only the leading nonzero power in $\lambda$ in the
last line, which, in effect, enforces the multipole expansion.  Here, we
are using the scalings with $\lambda$ of $\ell$ and $p$ that are given
in Eqs.~(\ref{eq:soft-ell-power}) and (\ref{eq:p-collinear-power}),
respectively.  We see that the expression in
Eq.~(\ref{eq:diagram-a}) has an additional power $\lambda^2$ relative
    to the contribution in Eq.~(\ref{eq:rad-Feyn-all}). That is, it is
power suppressed.

The diagram of Fig.~\ref{fig:AA-LO}(d) gives
\begin{eqnarray}
\label{eq:diagram-f}
&&
A_{\textrm{(d),Feynman}}(\ell^-)
\nonumber \\
&=&
(T^e)^{bc}
(T^e)^{da}
\left\{
g_s
\left(
{n}_\mu
+
\gamma_{\perp\mu}
\frac{\slashed{p}_\perp}{\bar{n}\cdot p}
\right)
P_n
\Pi_{^3S_1^{[1]}}^{cd}
(ig_s\gamma^\mu)
\frac{i}{2\slashed{p}-\slashed{\ell}-m+i\varepsilon}
P_{\bar{n}}
\right\}^{\beta\alpha}
\frac{-i}{\left(p-\ell\right)^2+i\varepsilon}
\nonumber \\
&=&
-
\frac{ig_s^2
C_F\delta^{ba}
}{\sqrt{2N_c}mp^+}
\frac{1}{(-\ell^-+i\varepsilon)}
\frac{
-m^2
\left(\frac{\slashed{\bar{n}}\slashed{n}\slashed{\epsilon}^*}{4}\right)^{\beta\alpha}
+m\left(\frac{\slashed{\bar{n}}\slashed{n}\slashed{\epsilon}^*
\slashed{\ell}_\perp}
{4}\right)^{\beta\alpha}
}{2p^+(-\ell^-+i\varepsilon)},
\end{eqnarray}
where we have kept only the leading nonzero power in $\lambda$ in the
last line. Again, we are using the scalings with $\lambda$ of $\ell$ and
$p$ that are given in Eqs.~(\ref{eq:soft-ell-power}) and
(\ref{eq:p-collinear-power}), respectively.  The expression in
Eq.~(\ref{eq:diagram-f}) has an additional power $\lambda^2$ relative
    to  the contribution in Eq.~(\ref{eq:rad-Feyn-all}). That is, it
is power suppressed.


\end{document}